\documentclass[nofootinbib]{revtex4-2}
\pdfoutput=1
\usepackage{graphicx}
\usepackage{dcolumn}
\usepackage{bm}
\usepackage{subfigure}
\usepackage{ulem}
\usepackage[table]{xcolor}
\usepackage{amsmath}
\usepackage{fancyvrb}
\usepackage[Lenny]{fncychap}
\usepackage{mathrsfs}
\usepackage{graphicx}  
\graphicspath{{img/}}
\usepackage{dcolumn}   
\usepackage{bm}        
\usepackage{amsmath,amssymb}
\usepackage{hyperref}
\usepackage{color}
\definecolor{purple}{rgb}{0.58,0.0,0.83}
\usepackage{caption}
\usepackage{tabularx}
\usepackage{comment}

\newcommand{\Like}{\mathcal{L}}

\def\beq{\begin{equation}}
\def\eeq{\end{equation}}

\definecolor{mycolor}{rgb}{0.3, 0.0, 0.3}

\definecolor{owngreen}{rgb}{0.0, 0.5, 0.0}

\newcommand{\changes}[1]{\textcolor{black}{#1}}
\begin{document}

\title{Coupled Multi Scalar Field Dark Energy} 

\author{J. Alberto V\'azquez$^1$}
\email{javazquez@icf.unam.mx}
\author{David Tamayo$^2$}
\email{david.tr@piedrasnegras.tecnm.mx}
\author{Gabriela Garcia-Arroyo$^1$}
\email{arroyo@icf.unam.mx}
\author{Isidro G\'omez-Vargas$^1$}
\email{igomez@icf.unam.mx}
\author{Israel Quiros$^3$}
\email{iquiros@fisica.ugto.mx}
\author{Anjan A. Sen$^4$ }
\email{aasen@jmi.ac.in}

\affiliation{$^1$Instituto de Ciencias Físicas, Universidad Nacional Autónoma de México,
62210, Cuernavaca, Morelos, México.}

\affiliation{$^2$Instituto Tecnológico de Piedras Negras, Calle Instituto Tecnológico 310 C.P. 26080, Piedras Negras, Mexico.}

\affiliation{$^3$Departamento Ingenier\'ia Civil, Divisi\'on de Ingenier\'ia, Universidad de Guanajuato, Guanajuato, C.P. 36000, M\'exico.}

\affiliation{$^4$Centre for Theoretical Physics, Jamia Millia Islamia, New Delhi-110025, India.}

\begin{abstract}
The main aim of this paper is to present the multi scalar field components as candidates to be the 
dark energy of the universe and their observational constraints. 
We start with the canonical Quintessence and Phantom fields with quadratic potentials 
and show that a more complex model should bear in mind to satisfy current cosmological observations. 
Then we present some implications for a combination of two fields, named as Quintom models. 
We consider two types of models, one as the sum of the quintessence and phantom potentials and other 
including an interacting term between fields. We find that adding one extra degree of freedom, by the interacting 
term, the dynamics enriches considerably and could lead to an improvement in the fit of {
$-2\ln\Delta \Like_{\rm max}= 5.19$}, compared to $\Lambda$CDM. The resultant effective equation of state 
is now able to cross the phantom divide line, and in several cases present an oscillatory or 
discontinuous behavior, depending on the interaction value. The parameter constraints of the scalar 
field models (quintessence, phantom, quintom and interacting quintom) were performed using Cosmic 
Chronometers, Supernovae Ia and Baryon Acoustic Oscillations data; and the Log-Bayes
factors were computed to compare the performance of the models.
We show that single scalar fields may face serious troubles and hence the necessity of a more complex 
models, i.e. multiple fields.

\end{abstract}
\maketitle

\section{Introduction}

The current accelerated cosmic expansion is supported by multiple observations such as the Type Ia 
supernovae (SNIa), the distribution of the Large Scale Structure (LSS), the Cosmic Microwave Background 
anisotropies (CMB), and the Baryon Acoustic Oscillation peaks (BAO); see \cite{Huterer:2017buf, Abdalla:2022yfr} 
and references therein.
These measurements may be an indication of a negative-pressure contribution, to the total energy density of 
the universe, as being the responsible to drive the accelerated expansion, commonly known as dark energy.
Among numerous candidates that play the role of the dark energy, the simplest and well-known 
is the cosmological constant term ($\Lambda$) introduced to the Einstein field equations, whose main feature 
lays down on having a constant energy density in time and be uniformly distributed in space.
The cosmological constant, along with the Cold Dark Matter, are the key elements that conform the standard 
cosmological model or $\Lambda$CDM. Some of the essential properties to understand the nature of the 
dark energy are encapsulated into its  equation of state (EoS); for a barotropic perfect fluid, it is 
defined as the ratio of the pressure over its energy density $w=p/\rho$.
In particular the $\Lambda$CDM model, having a dark energy EoS $w = -1$, describes very accurately most of 
the observational data, however in recent studies it seems to display a tendency in favor of a time evolving 
dark energy EoS $w(z)$ 
\cite{Copeland:2006wr, Zhao:2017cud, SolaPeracaula:2018wwm, Wang:2019ufm, Chudaykin:2020ghx, Yang:2021flj}.
Therefore several dark energy models with departures from the basic standard model have been introduced to 
take into account the evolution of $w(z)$, in addition to other properties \cite{DiValentino:2021izs},
%
for instance, the single scalar fields, as they have already been considered in cosmology to explain different 
phenomena, such as inflation, dark matter, modified gravity and also are excellent candidates for modeling the 
variable dark energy EoS \cite{Copeland:2006wr, Bahamonde:2017ize, Quiros:2019ktw}.  
Two well-known single scalar fields (one degree of freedom) have been extensively investigated 
for modeling dark energy are quintessence \cite{Tsujikawa:2013fta} and phantom \cite{Ludwick:2017tox}. 
The Lagrangian of both of them has a kinetic term and an associated potential, but the key distinction lays 
in the sign of their kinetic terms. For quintessence, its kinetic energy density is positive, while for phantom  
it is considered as negative. This slight difference results in distinct branches of values for their associated 
EoS parameter, i.e. for phantom $w < -1$ and for quintessence $-1 < w < 1$.
Furthermore, the phantom divide line (PDL), defined as $w(z)=-1$, separates the phantom-energy-like behavior 
with $w<-1$ from the quintessence-like behavior with $w>-1$, and the existence of a no-go theorem shows that 
in order to cross the PDL it is required at least two degrees of freedom for the models (in four dimensions) 
involving ideal gases or scalar fields\footnote{Bear in mind that for extra-dimensional models of dark energy, 
a single scalar field is able to cross the PDL \cite{Chimento:2006ac}.}, and this is where simple scalar 
fields may fail \cite{Cai:2009zp}. 
It is important to note that by using model independent techniques or non-parametric approaches i.e. Artificial 
Neural Networks, Gaussian processes or Nodal reconstructions, multiple studies have shown a preference for 
a crossing of the PDL in the dark energy EoS parameter 
\cite{Escamilla:2021uoj,Vazquez:2012cen, Hee:2016nho, Zhao:2017cud, Wang:2018fng, Wang:2019ufm}, 
which could also alleviate the Hubble tension and some inconsistencies among datasets, 
i.e. Ly-$\alpha$ and galaxy BAO data \cite{Panpanich:2019fxq}. Following this line of research, different 
groups reconstructed the general form of $w(z)$ and have converged to similar shapes 
\cite{Zhao:2017cud, Wang:2018fng, Dai:2018zwv}.
If this trend keeps on going in the forthcoming experiments, single canonical scalar field models may face 
serious troubles and hence more complex theories or more than a single field would be needed to explain 
this important feature. 
\\

In order to model the richness of the evolution of $w(z)$ we need more than the usual quintessence/phantom 
dark energy and thus invoke more elaborated models. In \cite{Hu:2004kh} the authors have constructed an 
EoS that crosses the PDL and it is based on a two field model.
Within the scalar field scenario, a scalar field dark energy model with an EoS parameter that traverses 
the PDL during its evolution, was firstly advocated and named as quintom dark energy in \cite{Feng:2004ad}.
Quintom is the next natural step for quintessence/phantom. This is a model that joins the quintessence and 
phantom fields (the reason behind its name) by considering both the positive and negative kinetic terms 
along with the potentials. 
Given that the null energy condition (NEC) is violated by the phantom degree of freedom~\cite{Cai:2009zp}, 
a quintom scenario is primarily designed for models with the NEC violation. 
The NEC violating degree of freedom leads to a quantum instability so that the fundamental origin of the 
quintom field poses a challenge for the theoreticians. 
Nevertheless, if viewed just as an effective (classical) cosmological field, quintom models represent an interesting set up
for PDL crossing dark energy models.
Quintom dark energy \cite{yifucai:2010pr} has been studied from various perspectives: theoretical aspects \cite{Qiu:2010ux}, 
observational constraints \cite{Zhang:2018zwu}, dynamical system approach 
\cite{Lazkoz:2007mx, Leon:2018lnd, Mishra:2018dzq, Panpanich:2019fxq}, non-minimal coupling 
\cite{Marciu:2016aqq}, and quantum cosmology \cite{Socorro:2013eba, Dutta:2021kjg}. 
Quintom models encompass interesting features of both quintessence and phantom: phantom dark energy has 
to be more fine tuned than quintessence in the early universe to serve as dark energy today, since its 
energy density increases with the expansion of the universe; meanwhile the quintom model mitigates the need 
for excessive fine-tuning by preserving -before the phantom domination- the tracking behavior of quintessence. 
Other research areas have also included similar ideas where two or more fields are present, for instance 
two scalar fields, or one field and its excited states, as being the dark matter 
\cite{Vazquez:2020ani, benisty2019two, Navarro-Boullosa:2023bya, Tellez-Tovar:2021mge}, a combination of 
the inflaton and the scalar field dark matter
\cite{padilla2019scalar}, the presence of the inflaton and the curvaton field \cite{Benisty:2018fja}, 
two scalar fields to account for inflation \cite{Bamba:2015uxa, vazquez2018inflationary}, interactions between 
dark energy and dark matter \cite{bertolami2012two} or the axiverse model 
\cite{arvanitaki2010string, mehta2021superradiance, cicoli2022fuzzy} (see also \cite{gutierrez2022scalar}). 
\\

On the other hand, many investigations of model-independent techniques, along with current cosmological observations, 
have not obtained just a dynamical dark energy EoS but some of these results present wavering behaviors 
for the EoS, starting at $z=0$ with $w<-1$, crossing the PDL, then presenting a maximum and crossing 
back the PDL, even in multiple occasions. This type of model independent behavior suggests the necessity 
to include two or more fields, or the considerations of more complex potentials or even the coupling between 
these fields. 
A quintom model with an oscillating EoS was considered first in \cite{Feng:2004ff}, starting from the idea 
of a wavering behavior. In this paper we will extend the work of single fields in \cite{Vazquez:2020ani}, 
to multiple fields and revisit the quintom model with the addition of an interacting term, which may produce 
the oscillatory behavior and for some particular types of potentials can be justified by a symmetry group.\\


The paper is organized as follows: in section \ref{Quintom dark energy model} we summarize the main 
characteristics of the multi scalar field dark energy models; in section \ref{quintom model with interacting term} 
a novel quintom model with an interacting term is presented and described its main characteristics; 
in \ref{code an observations} we present the datasets and statistical techniques used to constrain the 
cosmological parameters associated to the model; in section \ref{results} we present the main results; 
and finally, in section \ref{conclusions} we summarize our results and provide some comments and conclusions.

\section{Multi Scalar field dark energy model}\label{Quintom dark energy model}

In the context of four dimensional spacetime, it is not feasible for a single scalar field to serve as a viable dark 
energy candidate for modeling the crossing of the phantom barrier. Therefore, it is necessary to introduce 
extra degrees of freedom, or to introduce the non-minimal couplings, or to modify the Einstein gravity.
Adding extra degrees of freedom to the single scalar field dark energy requires the simultaneous consideration 
of more fields, as in the constructed quintom model which contains one canonical quintessence $\phi_1$ and 
one phantom $\phi_2$, and therefore the dark energy is attributed to their combination.
The action of a cosmological model that incorporates multiple real scalar fields $\phi_i$, is given by
\begin{equation}
    S = \int d^4x\sqrt{-g}\left[\frac{R}{2\kappa^2} 
        + \frac12 \sum_i \epsilon_i \partial^\nu \phi_i \partial_\nu \phi_i  
        -V(\vec \phi)  + \mathcal{L}_M\right],
\end{equation}
where $\kappa^2 = 8\pi G$ is the gravitational coupling and the term $\mathcal{L}_M$ accounts for the 
remaining cosmological components of the universe (dark matter, baryons, radiation, etc.).
The index $i$ represents the number of fields with a total associated potential $V(\vec \phi)=V(\phi_1,\cdots,\phi_i)$; 
and the $\epsilon_i$ parameter is restricted to take either one of the two values $\epsilon_i=\{+1, -1 \}$ in 
order to account for the distinction between quintessence ($+1$) and  phantom ($-1$) fields respectively.

%
%
%
%
Considering a flat Friedman-Robertson-Walker space-time,  the Friedmann equations are thus
\begin{eqnarray}
    H^2 &=& \frac{\kappa^2}{3}(\rho_Q +\rho_M), \label{eq:Friedman1}\\
    \dot{H} &=& -\frac{\kappa^2}{2}(\rho_Q +p_Q +\rho_M+p_M), \label{eq:Friedman2}
\end{eqnarray}
where $H$ represents the Hubble parameter and an over-dot denotes differentiation with respect to
cosmic time. The standard energy density components $\rho_M = \sum \rho_j$, are assumed to be perfect fluids 
and have a barotropic EoS of the form $w_j= p_j/ \rho_j$. Hence, the standard energy conservation equation for 
each one reads as
\begin{eqnarray}
    \dot{\rho}_j + 3H (1+\omega_j) \rho_j = 0.
\end{eqnarray}
In the case of pressureless matter we have $w_j =0$, whereas for the relativistic particles $w_j = 1/3$. 
For the multi-fields, the associated total energy density and pressure are given by
\begin{equation}\label{multi_field}
    \rho_Q = \frac{1}{2}\sum_i \epsilon_i \dot{\phi_i}^2 +V(\vec \phi), \qquad
    p_Q = \frac{1}{2}\sum_i \epsilon_i \dot{\phi_i}^2  -V(\vec \phi),
\end{equation}
and the EoS of the combined fields, i.e, the total effective EoS is then
\beq \label{eq:eos}
	w_Q = \frac{\sum_i \epsilon_i \dot{\phi_i}^2 - 2V(\vec \phi)}{\sum_i \epsilon_i \dot{\phi_i}^2 
    + 2V(\vec \phi)} \, ,
\eeq
whose value can only be determined from the evolution of the fields themselves. The dynamics of the scalar 
fields is determined by solving, the following coupled Klein-Gordon equation
\beq \label{eq:KGs}
\sum_i \left[\dot \phi_i\left(\epsilon_i\ddot\phi_i + 3H\epsilon_i\dot\phi_i 
    +\frac{\partial V(\vec \phi)}{\partial\phi_i}\right)\right]  = 0.
\eeq
For the particular case of $\phi_1 = \phi$ and $\phi_2 = \psi$,  $\epsilon_1=1$ and $\epsilon_2=-1$, then  
 $V =V(\phi, \psi)$, the last expression becomes:
\beq \label{eq:KG general}
\dot \phi\left(\ddot\phi + 3H\dot\phi +\frac{\partial V(\phi, \psi)}{\partial\phi}\right) 
    -\dot\psi \left(\ddot\psi + 3H\dot\psi -\frac{\partial V(\phi,\psi)}{\partial\psi}\right)  = 0.
\eeq
%
%
In general, this equation does not enforce a split into two coupled Klein Gordon equations, however 
this specific splitting represents a special case where the overall equation is satisfied
\begin{eqnarray}
\ddot\phi + 3H\dot\phi +\frac{\partial V(\phi, \psi)}{\partial\phi}=0,\\
\ddot\psi + 3H\dot\psi -\frac{\partial V(\phi,\psi)}{\partial\psi} = 0.
\end{eqnarray}

Clearly the quintom model,  $\phi_1 = \phi, \phi_2 = \psi$ $(\epsilon_1=1 ,\epsilon_2=-1)$, boils 
down into quintessence when the phantom field is null $\psi=0$, and conversely into phantom when $\phi=0$. 
In general, $\phi$ will evolve towards the local minima of the potential, whereas $\psi$ towards the local 
maxima; such different behaviors arise because of the signs in the Klein-Gordon equations, inherited from 
the signs of the kinetic energy terms in the action. 
\\

\begin{figure}[t!]
\captionsetup{justification=raggedright,singlelinecheck=false,font=small}
\begin{center}
\makebox[10cm][c]{
\includegraphics[trim = 2mm  0mm 2mm 2mm, clip, width=5.cm, height=9cm]{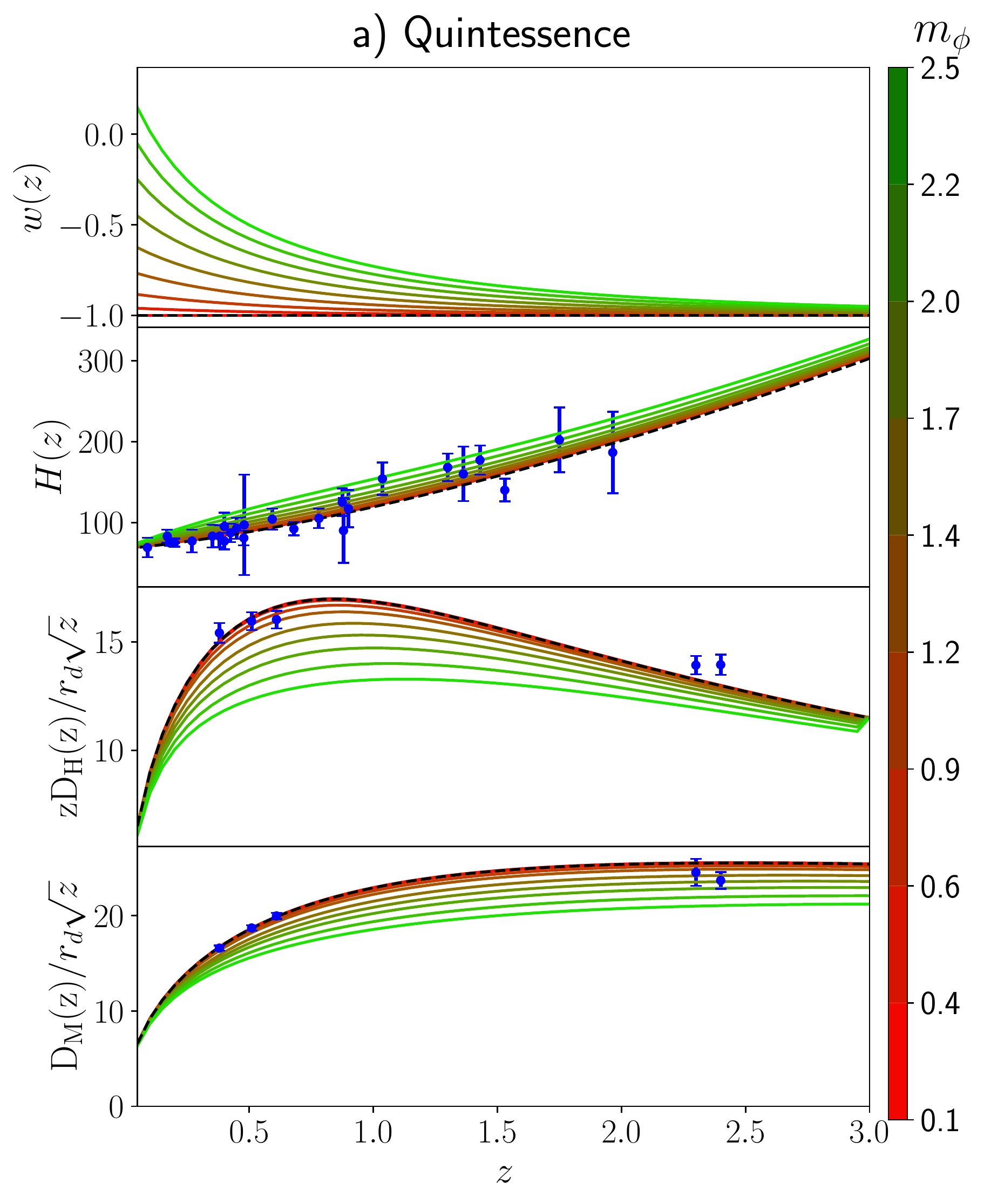} 
\includegraphics[trim = 2mm  0mm 2mm 2mm, clip, width=4.5cm, height=9cm]{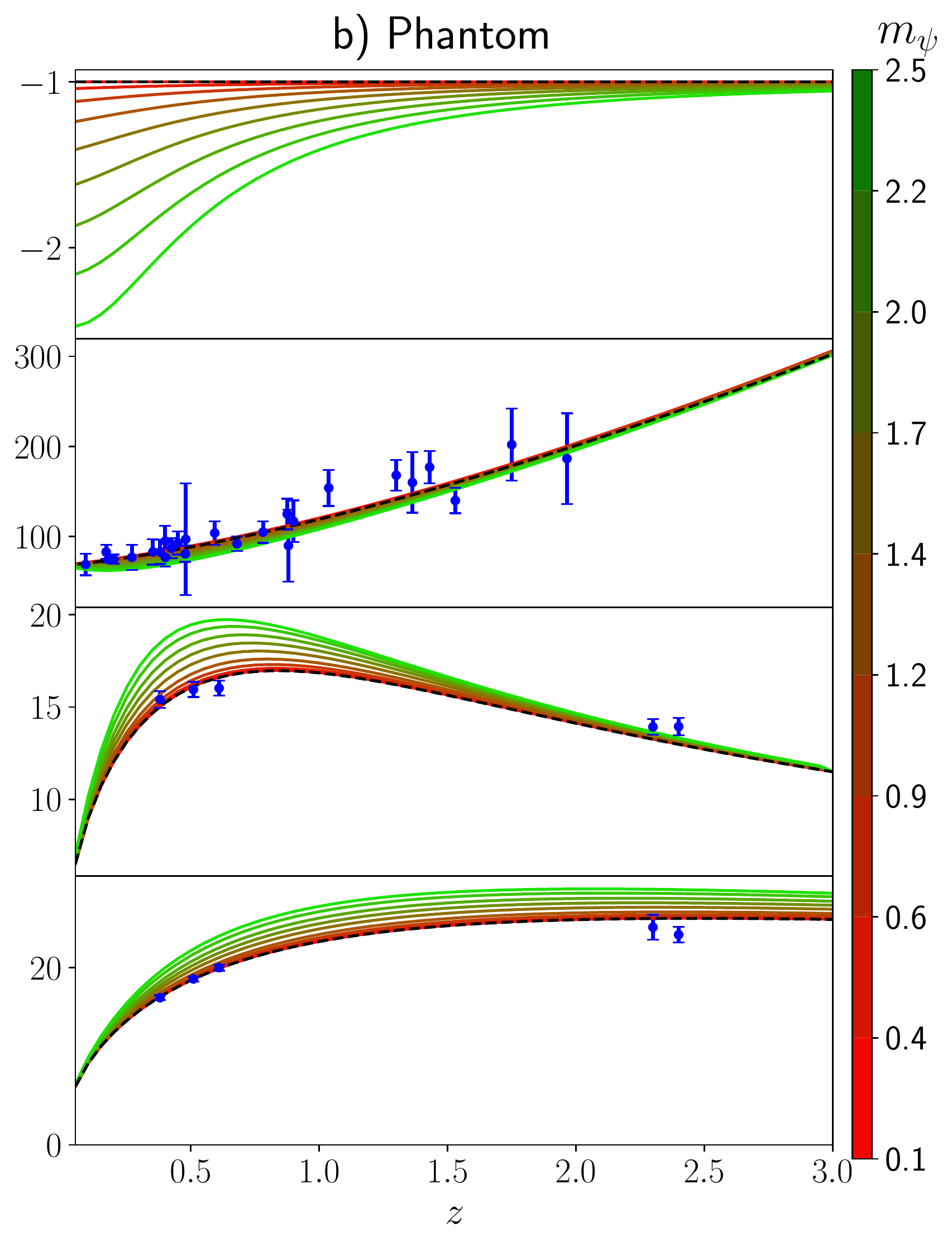} 
\includegraphics[trim = 2mm  2mm -4mm 2mm, clip, width=4.5cm, height=9cm]{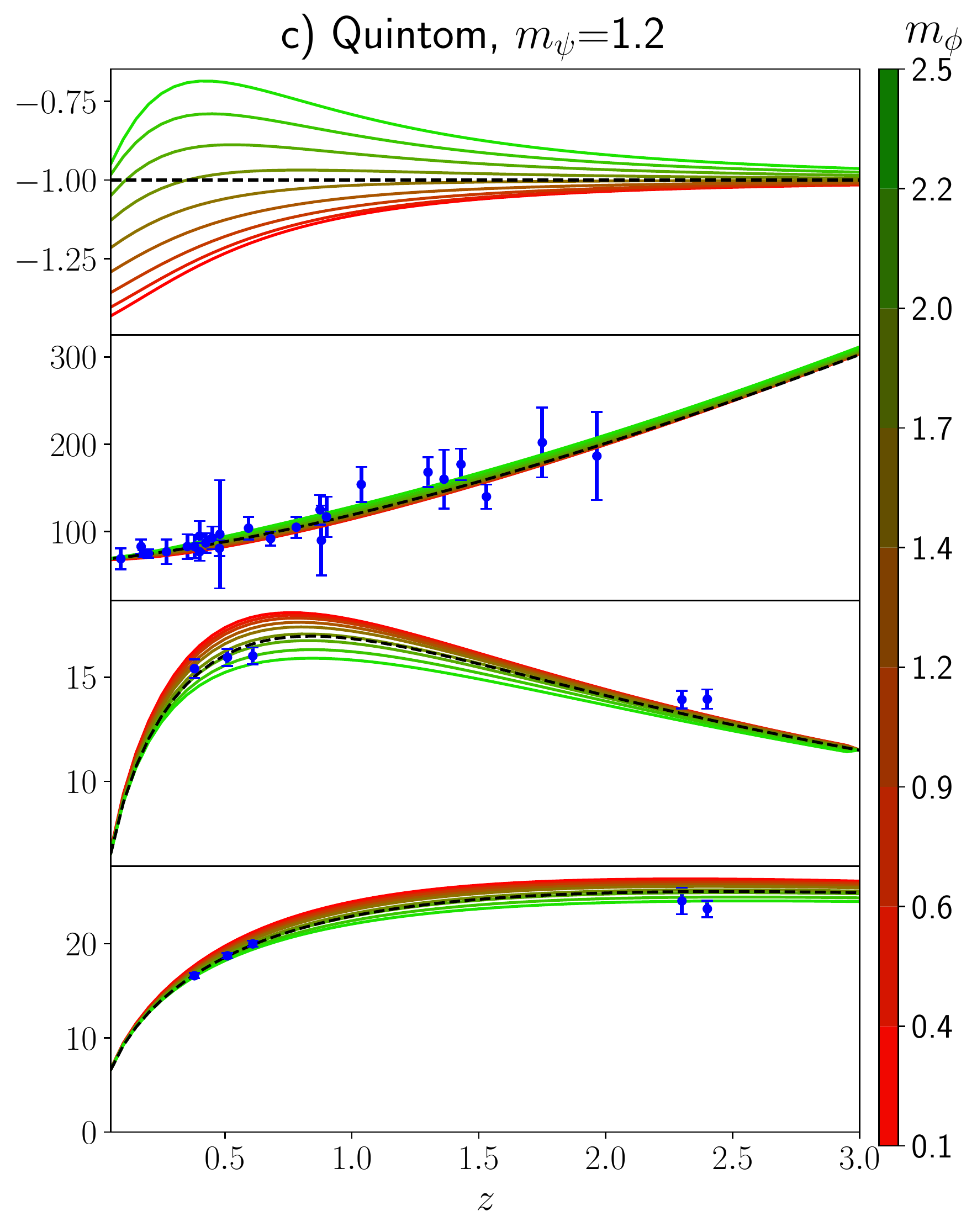} 
\includegraphics[trim = 0mm  1mm -2mm 2mm, clip, width=5.0cm, height=9cm]{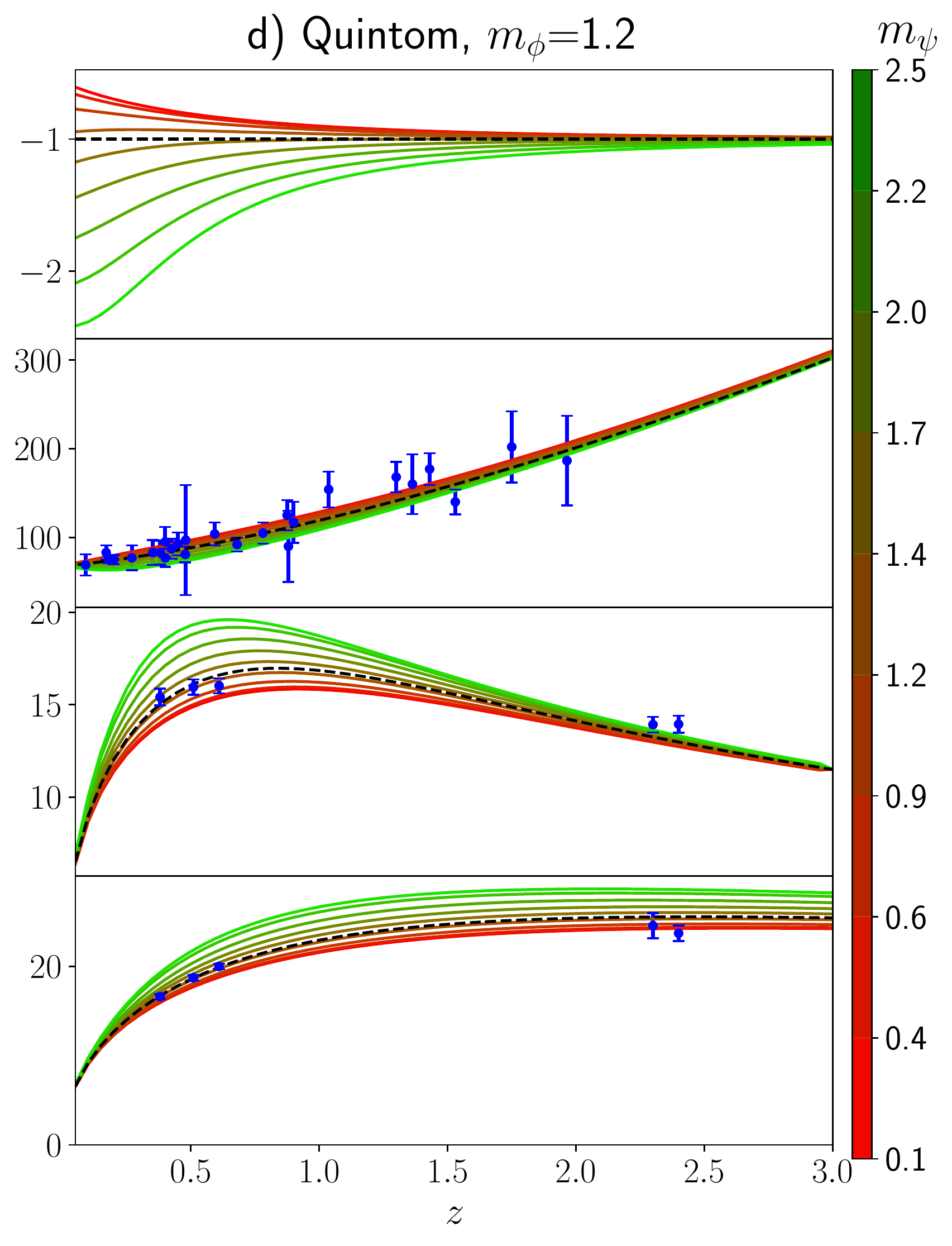} 
}
\end{center}
\caption[]{From left to right: quintessence, phantom and quintom cosmologies with potentials 
$V(\phi) = \frac{1}{2}m_{\phi}^2\phi^2$, $V(\psi) = \frac{1}{2}m_{\psi}^2\psi^2$ and $V(\phi,\psi) 
= \frac{1}{2}m_{\phi}^2\phi^2+ \frac{1}{2}m_{\psi}^2\psi^2$, respectively.
The first row displays the EoS $w(z)$, the second the Hubble function $H(z)$, the third the Hubble 
distance $D_H$ and the fourth the comoving angular distance $D_M$.
The color bar represents different values for the masses of the fields. The data plotted for $H(z)$ correspond 
to the cosmic chronometers \cite{jimenez2003constraints, simon2005constraints, stern2010cosmic, moresco2012new, 
zhang2014four, moresco2015raising, moresco20166, ratsimbazafy2017age}; for the $D_H$ and $D_M$ panels we use 
the BAO Galaxy consensus ($z \sim 0.5$) \cite{alam2017clustering}, Ly-$\alpha$ DR14 auto-correlation ($z=2.34$)
\cite{de2019baryon} and cross-correlation ($z=2.35$) \cite{blomqvist2019baryon}.
In all panles, the black dashed line describes the $\Lambda$CDM model.
}
\label{Fig:quint-phan}
\end{figure}

%
Finally, in order to determine the dynamics of the system, it is necessary to solve the conservation 
equations for the matter components and scalar fields.
Following \cite{Vazquez:2020ani},  the Klein-Gordon equations can be rewritten as a dynamical system 
that can be solved straightforward, where the initial conditions for each field have been set up right into 
the matter domination epoch, and we have assumed a thawing behavior for the multi fields. 
This implies that at early times, the kinetic terms of the quintom model vanish, and its equation of state, 
$w_Q$, begins at $-1$. Also, the initial scalar field density parameter $\Omega_{Q,{\rm ini}}$ is selected, 
through a shooting mechanism, such that its present value satisfies the Friedmann constraint, 
$\Omega_{Q, 0}+ \Omega_{M, 0}= 1$. 
To illustrate the general behavior of the quintessence, phantom and quintom models, in Figure 
\ref{Fig:quint-phan} we plot the following cosmological quantities: (from top to bottom) EoS $w(z)$, the 
Hubble function $H(z)$, the Hubble distance $D_H$ and the comoving angular distance $D_M$ along with 
several measurements (see the caption's figure). 
In all the panels, the $\Lambda$CDM model is represented by a black dashed line. As a proof of the concept, 
the potentials used in our analysis are as followed: for quintessence and phantom the quadratic potential, 
and for quintom the sum of the two aforementioned potentials\footnote{For an ample variety of potentials, 
refer to \cite{Vazquez:2020ani}}
\begin{eqnarray}
    \text{quintessence:} \quad V_{\phi} &=&  \frac{1}{2}m_{\phi}^2\phi^2, \label{eq:Vphi}\\
    \text{phantom:} \quad V_{\psi} &=& \frac{1}{2}m_{\psi}^2\psi^2, \label{eq:Vpsi}\\
    \text{quintom:} \quad V_Q &=& \frac{1}{2}m_{\phi}^2\phi^2 +\frac{1}{2}m_{\psi}^2\psi^2.\label{eq:Vtot}
\end{eqnarray} 
Here $m_{\phi}$, $m_{\psi}$ are the masses of the fields; both of them plotted in units of [3$H_0$]. 
To show the main behavior of the inclusion of the fields, we varied only the masses and kept fix the 
rest of the cosmological parameters. Notice, that in general, for masses tending to null values we recover
the $\Lambda$CDM pattern.
In panels a) and b) of Figure \ref{Fig:quint-phan}, the quintessence model shows an EoS $w>-1$ meanwhile for 
phantom $w<-1$. Notice that, in both cases, as the masses of the fields increase, the EoS deviates 
farther from the cosmological constant line ($w=-1$) in the late--time regime, however in opposite directions. 
In general, all the quintessence $H(z)$ lines remain above the Hubble function of the $\Lambda$CDM model; this 
increment also yields to $D_H$ and $D_M$ to lay below the $\Lambda$CDM; for the phantom model occurs 
qualitatively the opposite. Both behaviors have been already studied in~\cite{Vazquez:2020ani}.
On the other hand, for the quintom model and without loosing generality, we considered two 
cases -- panels c) and d) of Figure \ref{Fig:quint-phan} --. In the first case we fix the phantom mass 
at $m_{\psi}=1.2$ and let the quintessence mass $m_{\phi}$ vary;  while in the second case, 
we reversed the scenario. 
For a fixed $m_{\psi}$, panel c), the PDL crossing occurs when $m_\phi \gtrsim m_\psi$,  and depending 
on the ratio $m_\phi/ m_\psi$  the dark energy EoS presents a maximum which becomes more pronounced 
as $m_\phi/ m_\psi$ increases; on the other hand, for a fixed $m_{\phi}$ the PDL crossing occurs less 
pronounced and the EoS may exhibit a minimum depending on the combination of $m_\psi/ m_\phi$
\footnote{In this work we focus on these type of models, named as Quintom-A field, which main feature 
is that at late times the Phantom dominates ($w<-1$) where at early times it does the Quintessence ($w>-1$). 
However, it would be interesting to perform a similar analysis to its dual, a Quintom-B field, that 
mirrors the EoS behavior along the PDL axis \cite{Cai:2006dm}.}. 
Finally, in these last two cases, there is a mixed behavior for $H(z)$, $D_H$ and $D_M$, to stay above or 
below to the $\Lambda$CDM observables, depending on the mass-parameter combination.
In all cases the EoS converges to $w=-1$ at high redshift, by construction.

\section{Quintom model with interacting term}\label{quintom model with interacting term}

The previous section presented a quintom model with a non direct interacting term in the scalar field potential, 
i.e., it can be split up into two independent functions  of the fields $V(\phi, \psi)= V(\phi)+ V(\psi)$, 
therefore the Klein-Gordon eqs. \eqref{eq:KGs} are coupled only through the Friedmann equation. 
Now, a step further is to consider a scalar potential with an interaction term. {For a renormalizable model, a general form of the potential must include operators with dimension four or less, consisting of various powers of the scalar fields. A reasonable choice is to consider a potential which respects $Z_2$ symmetry, i.e., it is invariant under the following simultaneous transformations: $\phi\rightarrow-\phi$, $\psi\rightarrow-\psi$ \cite{xiong:2008plb}. A potential containing an interaction term between the fields $\phi$ and $\psi$ and with the above properties has the following form:}\footnote{This potential, can be derived, as well, from an inflaton-phantom unification protected by an internal $SO(1, 1)$ symmetry, with the two cosmological scalars appearing as the degrees of freedom of a sole fundamental representation \cite{Chepe2020, PerezLorenzana:2007qv}.}
\begin{equation}\label{eq:quintom potential}
    V(\phi, \psi) = \frac{1}{2}m_{\phi}^2\phi^2 +\frac{1}{2}m_{\psi}^2\psi^2 +\beta \phi^2\psi^2.
\end{equation}
{This is a particular case of the potential considered in Eq. (8) of Ref. \cite{xiong:2008plb} (see also Eq. (4) of Ref. \cite{Cai:2011bs}), when in the latter set the mass squared dimension's constant $\Lambda_0=0$ and make the identification $\lambda^2=\beta$. Although phantom and, also, quintom models, as the one we are considering here \eqref{eq:quintom potential}, suffer from a severe problem of quantum instability \cite{Carroll:2003st, Cline:2003gs}, it has been argued that, since we are considering a classical theory of gravity (GR), these fields should be considered as an appropriate effective description, no more.}

Due to the interaction term, $\beta \phi^2 \psi^2$, the equations of motion of the scalar fields are coupled, 
hence the term related to the potential in the Klein-Gordon equation of the field $\phi$ becomes $\frac{\partial V(\phi, \psi)}{\partial \phi} = m_{\phi}^2\phi + 2\beta \phi\psi^2 = f(\phi, \psi)$ which depends on both fields; the same happens for the $\psi$ field. We will refer to the quintom model with potential \eqref{eq:quintom potential} as the interacting quintom, or just `quintom$+\beta$'.\footnote{We based our analysis on this potential, however in the Appendix \ref{appendix_a} there are some other interesting alternatives to explore in future works.} 
{Although the above potential has been formerly considered to describe the qualitative behavior of a cyclic Universe \cite{xiong:2008plb, Cai:2011bs}, as long as we know, its observational consequences have not been previously investigated in detail. This is one of the reasons why we chose this specific quintom model for our study. The other reason, which has been already commented in the text, is that quintom models provide a feasible crossing of the phantom divide, a feature that seems to be favored by the observations.}

\begin{figure}[t!]
\begin{center}
\includegraphics[trim = 2mm  0mm 0mm 1mm, clip, width=5.8cm, height=4.2cm]{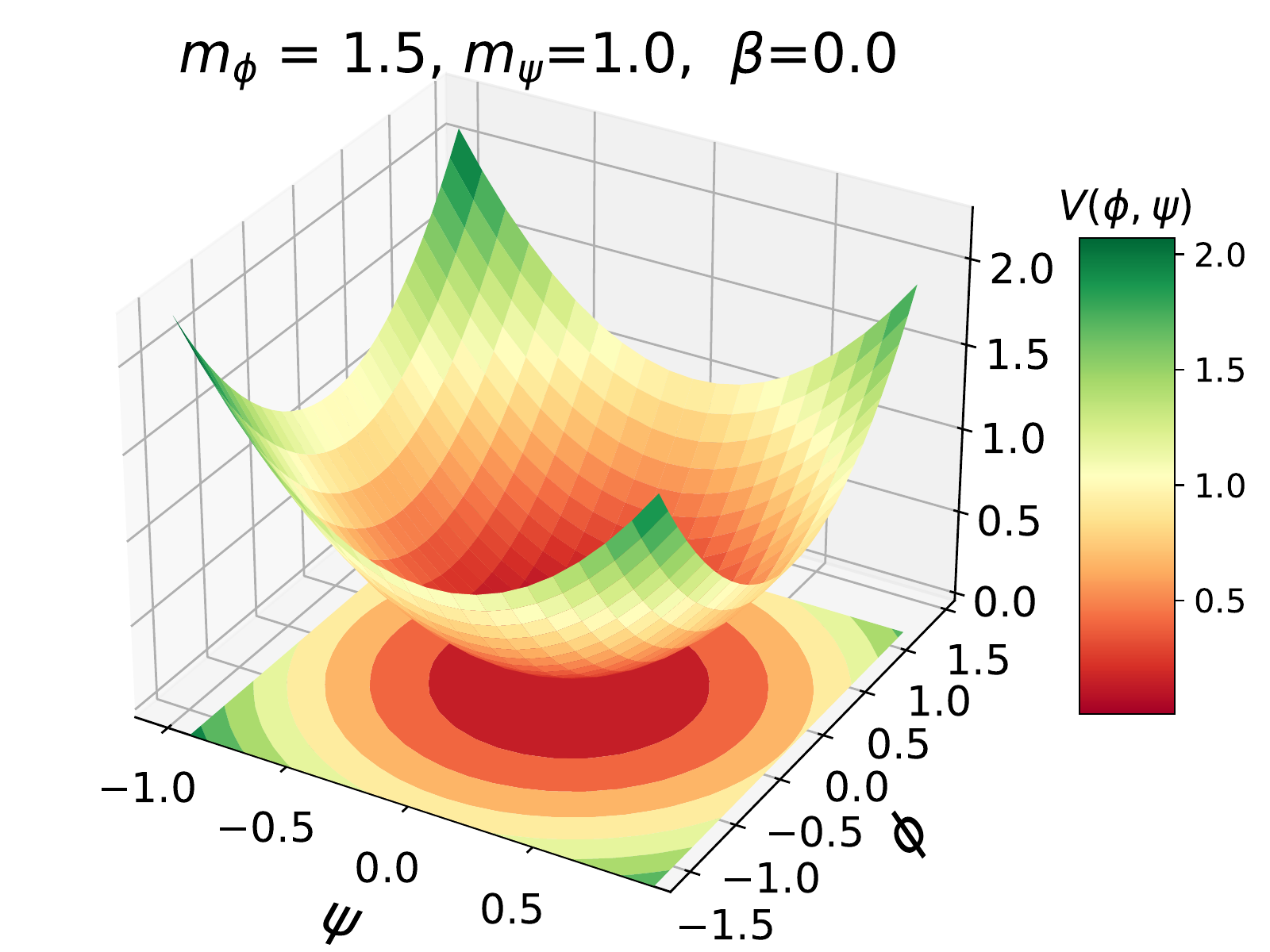} 
\includegraphics[trim = 2mm  0mm 0mm 1mm, clip, width=5.8cm, height=4.2cm]{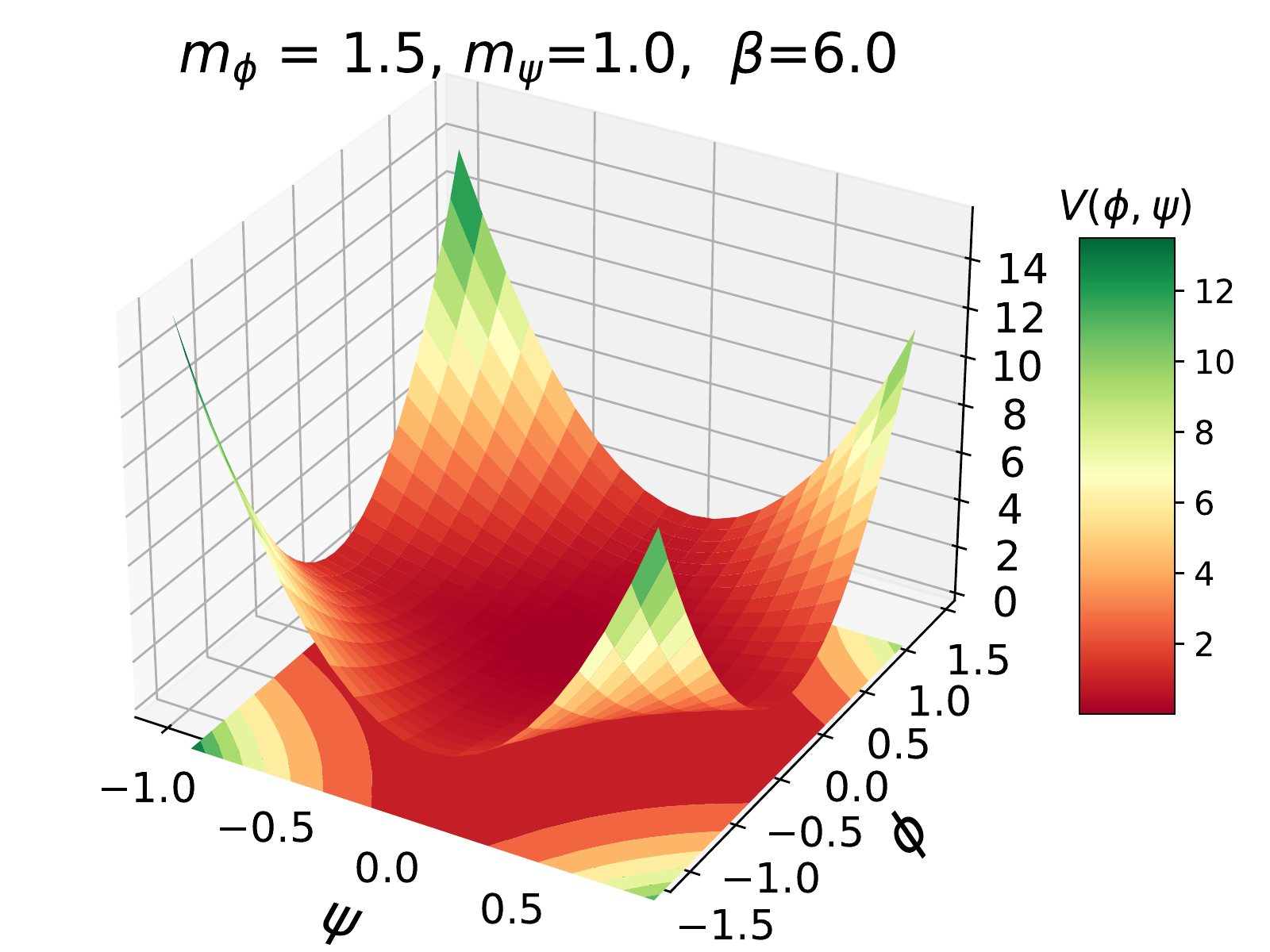}
\includegraphics[trim = 2mm  0mm 0mm 1mm, clip, width=5.8cm, height=4.2cm]{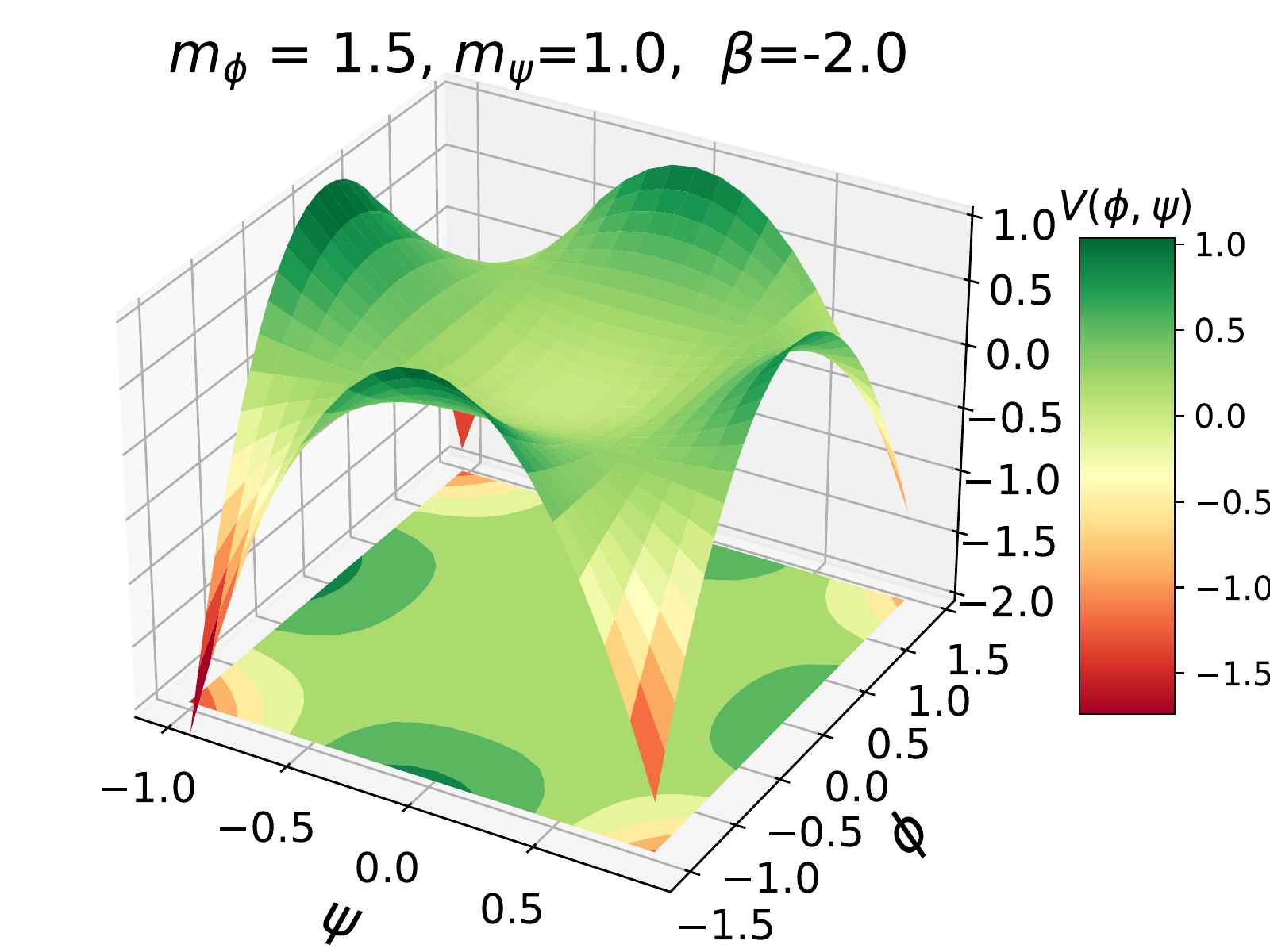}
\end{center}
\caption[]{Quintom potential (expression \eqref{eq:quintom potential}) for fixed values of the masses 
and positive, null and negative values of the coupling parameter $\beta$. The color bar indicates 
the values of the potential $V(\phi, \psi)$ ($z$-axis).}
\label{fig:potentials}
\end{figure}

Figure \ref{fig:potentials} displays the quintom$+\beta$ potential for fixed values of the masses 
$m_{\phi}=1.5$ and $m_{\psi}=1.0$, and three values of the interaction constant $\beta=0.0, 6.0, -2.0$. 
For the case $\beta=0$ we recover the results of the previous section. For positive $\beta$ there is a widen 
paraboloid with $V\geq 0$ whose minimum is at the region $\phi=\pm\frac{m_{\psi}}{m_{\phi}} \psi$, 
while in the negative $\beta$ case the potential may get to negative values (seen in the color bar of the figure), 
which can be problematic as they may produce a negative energy density. However, such behavior has been studied 
throughout negative dark energy models \cite{Akarsu:2019ygx, Calderon:2020hoc, Acquaviva:2021jov} and 
a sign switching cosmological constant $\Lambda$ \cite{Akarsu:2019hmw, Akarsu:2021fol, Sen:2021wld, Malekjani:2023dky}.
\\

\begin{figure}[t!]
\captionsetup{justification=raggedright,singlelinecheck=false,font=small}

\begin{center}
\makebox[10cm][c]{
\includegraphics[trim = 2mm  2mm 0mm 2mm, clip, width=5cm, height=9cm]{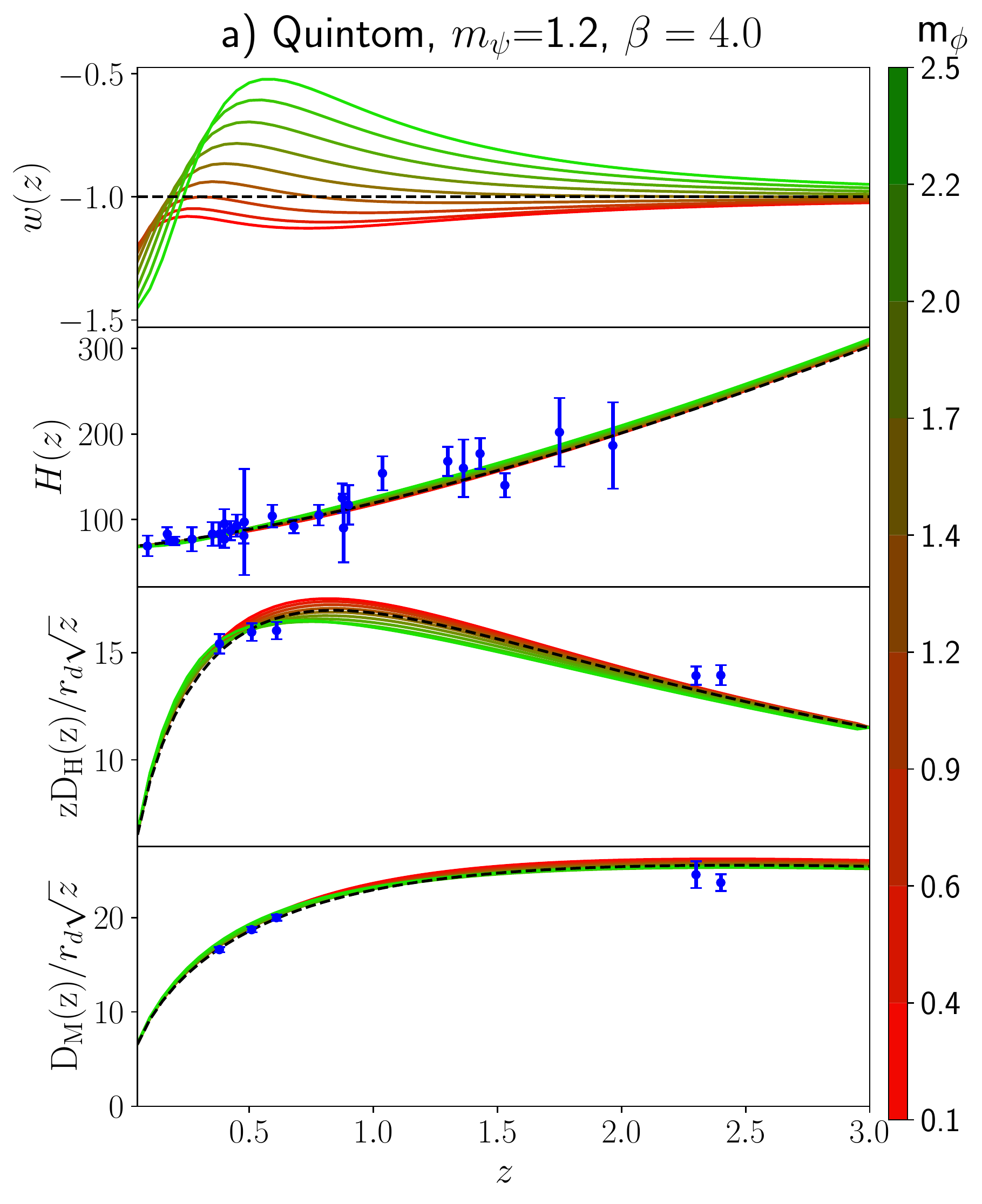}
\includegraphics[trim = 2mm  2mm 0mm 2mm, clip, width=4.5cm, height=9cm]{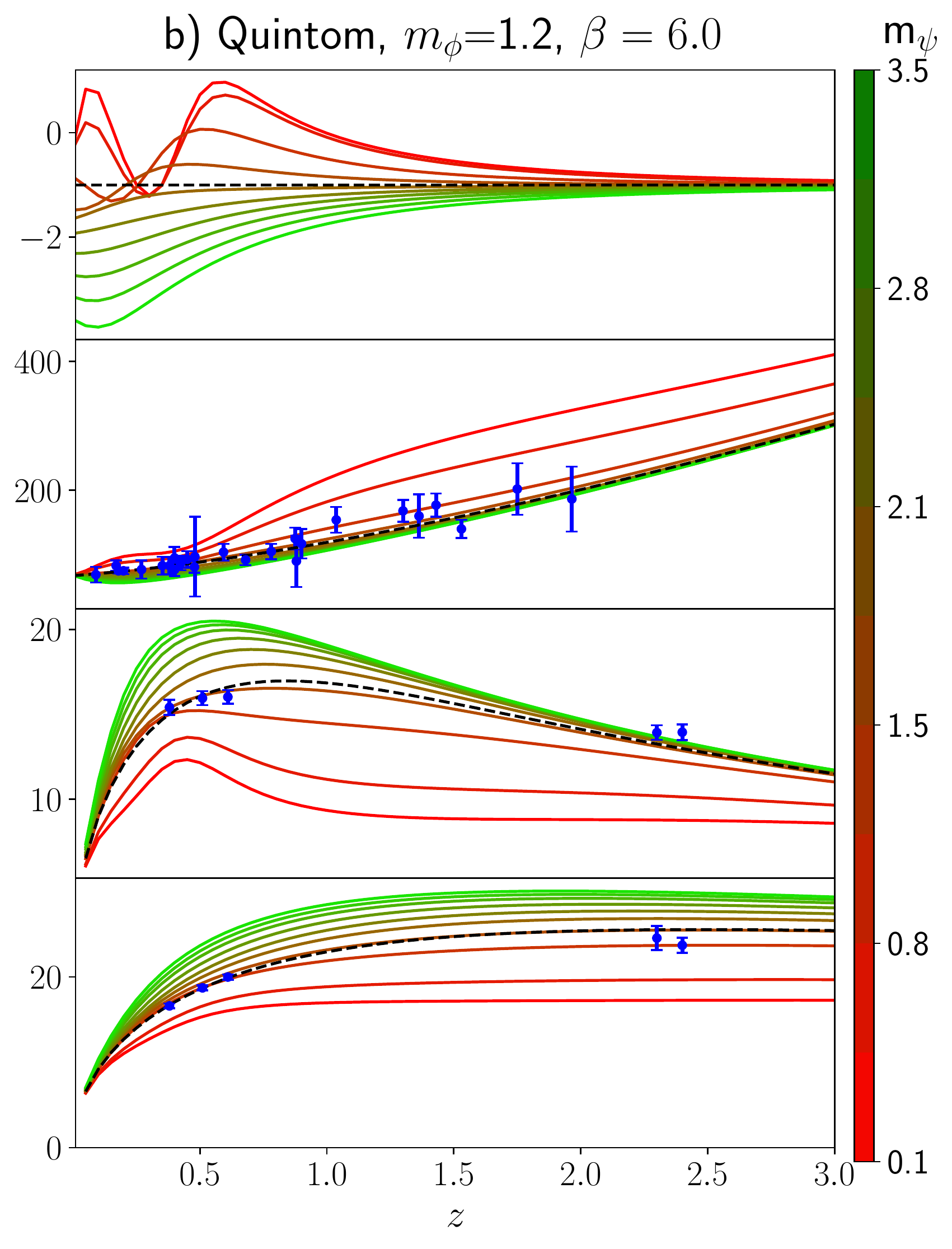}
\includegraphics[trim = 2mm  2mm 0mm 2mm, clip, width=4.5cm, height=9cm]{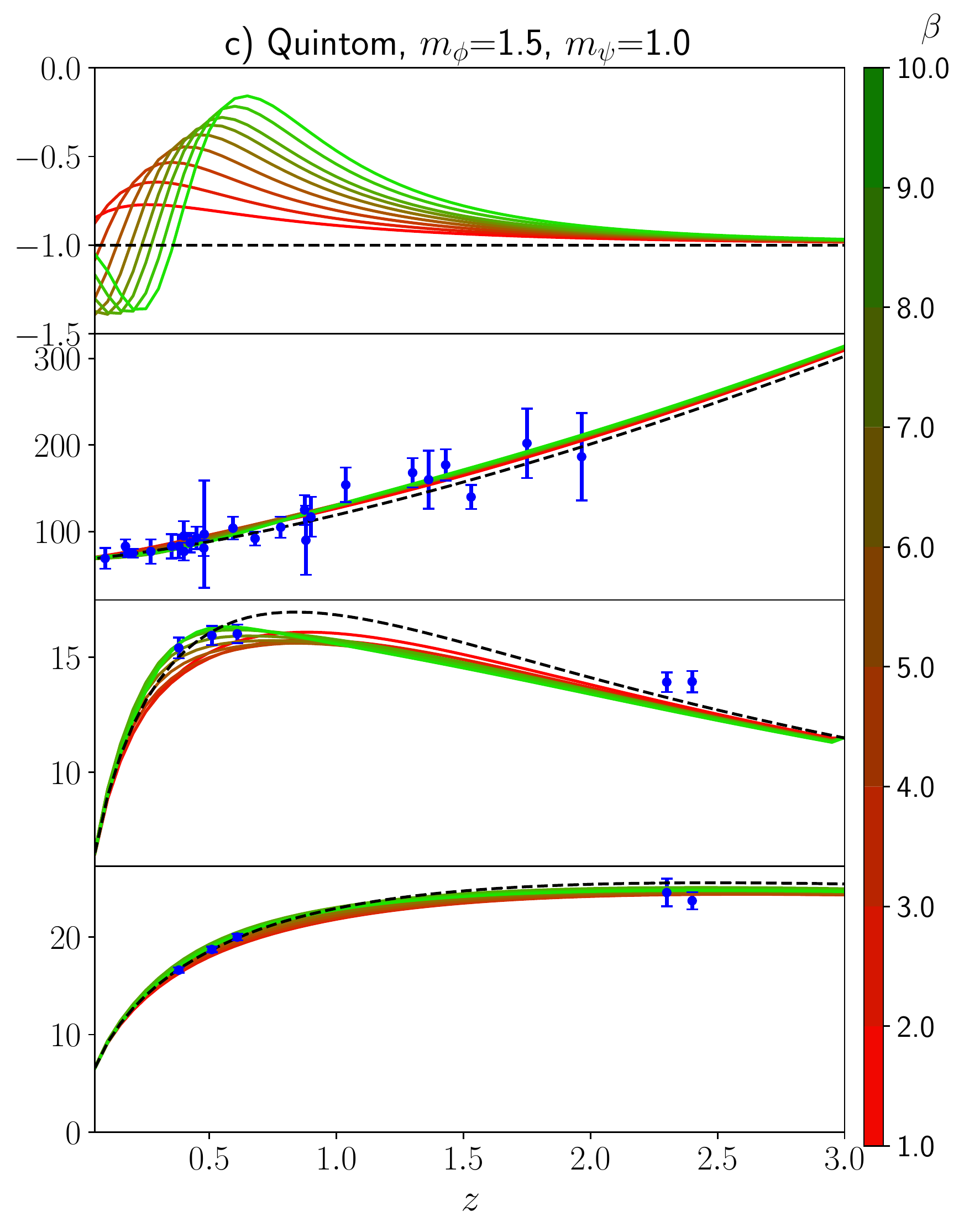}
\includegraphics[trim = 2mm  2mm 0mm 2mm, clip, width=5cm, height=9cm]{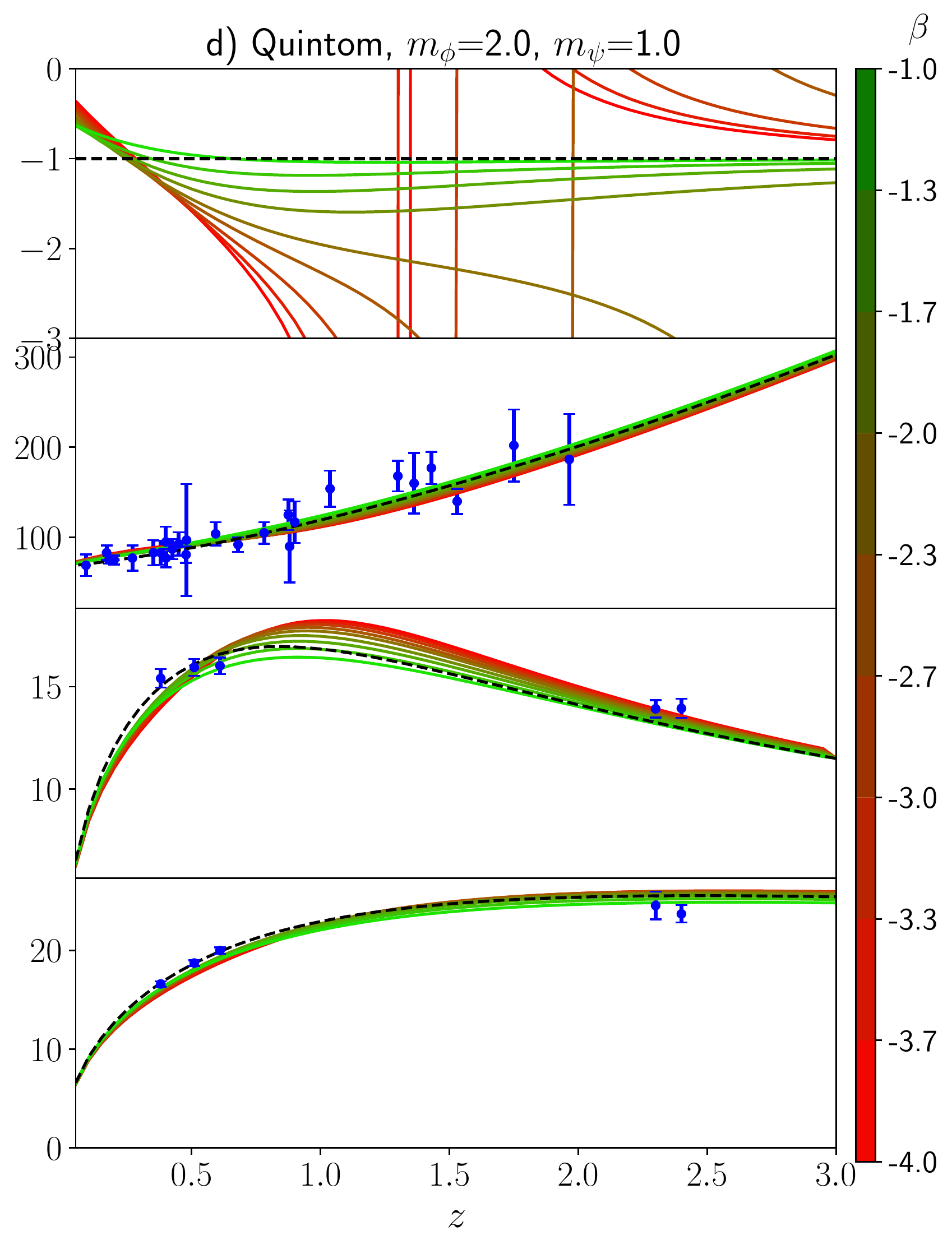}
}
\end{center}
\caption[]{Quintom model. Two of the three parameters ($m_{\phi}$, $m_{\psi}$, and $\beta$) are fixed 
while the remaining varies in a range of values. From left to right, in the first panel vary $m_{\phi}$, 
in the second $m_{\psi}$, in the third $\beta\geq0$ and in the fourth $\beta <0$.
The first row is the EoS $w(z)$, the second the Hubble function $H(z)$, the third the Hubble distance 
$D_H$ and the fourth the angular distance $D_M$. The plotted data points are the same as in the 
Figure \ref{Fig:quint-phan}.
The color bar represents different values for the masses of the fields or the $\beta$ parameter.}
\label{fig:mquin-mphan}
\end{figure}

Let us explore some cosmological implications of the quintom$+\beta$ model by  numerically solving the 
full dynamical system. Similarly to the previous cases, we compute the evolution of $w(z)$, $H(z)$, $D_H(z)$ 
and $D_M(z)$; their associated plots are shown in figure \ref{fig:mquin-mphan}.
In panel a) of this figure  (fixed $m_{\psi}=1.2$, $\beta =4.0$ and varying $m_{\phi}$), 
the EoS at $w(z=0)$ is on the phantom region, grows up and reaches a local/global maximum value. 
If the mass ratio condition $m_\phi/m_\psi < 1$ is satisfied (red lines), then the effective field evolves only in the 
phantom region, however for values $m_\phi\sim m_\psi$ it crosses to the quintessence regime, and in some combinations 
of the masses it is able to cross back into the phantom zone, therefore crossing twice the PDL. As the ratio $m_\phi/m_\psi$ 
increases (green lines), the maximum is shifted to larger redshifts and produces larger values of $w(z)$, 
to then remain in the quintessence region (the behavior of a Quintom-A field, as mentioned previously). 
An interesting point to note is that the quintom+$\beta$ model is able to traverse the Hubble $\Lambda$CDM line 
$H_{\Lambda{\rm CDM}}(z)$ (black dashed line). That is, if  $H_{\Lambda{\rm CDM},0}$ is larger than the Hubble 
parameter given by the $H_{Q,0}$ of quintom+$\beta$ model, $H_{\Lambda{\rm CDM},0}> H_{Q,0}$, then at some redshift 
it will occur that $H_{\Lambda{\rm CDM}}(z) < H_{Q}(z) $. This type of behavior provides flexibility to 
transverse the $\Lambda$CDM observables $D_H(z)$ and $D_M(z)$, contrary to the decoupled quintom model 
(see panel c) and d) of figure \ref{Fig:quint-phan}).

As the $\beta$ parameter increases (and depending on the combination of the masses) the oscillation is 
more pronounced and the crossing of the $\Lambda$CDM observables is more notorious,
see for instance panel b) of figure \ref{fig:mquin-mphan} ($m_{\phi}=1.2$, $\beta = 6.0$ and varying 
$m_{\psi}$). For $m_\psi/m_\phi <1$ (red lines), the EoS presents a wavering behavior 
and crosses the PDL multiple times. As the ratio $m_\psi/m_\phi$ increases, the number of oscillations are fewer 
with smaller amplitudes, until the evolution becomes a fully phantom behavior (green lines).
A similar oscillatory function of $w(z)$ has been found by using non-parametric reconstructions directly 
from observables \cite{Zhao:2017cud} and on phenomenological models that encompass these features 
\cite{Tamayo:2019gqj}.
Panel c) of figure \ref{fig:mquin-mphan} displays the outcomes for fixed mass values
$m_{\phi}=1.5$, $m_{\psi}=1$ and taking only positive values of $\beta$. As the $\beta$ parameter 
increases the wavering behavior enhances, and both the crossing of the PDL and the maximum value of 
the EoS shift to higher redshifts. 
\\

Finally, in panel d) of figure \ref{fig:mquin-mphan}  we show some peculiar behaviors when  
the negative case of $\beta$ is taken into account; here we use $m_{\phi}=2.0$ and $m_{\psi}=1$.
At the present time, $w(z)$ resides in the quintessence region, opposite to the $\beta \geq 0$ case, 
but in the past, it crosses the PDL and for high negative values of $\beta$ the EoS diverges, 
to then come back to the region quintessence $w(z)>-1$ at early times.
The presence of a pole in the EoS has been studied under several different physical circumstances 
in \cite{Sahni:2004fb, Tsujikawa:2008uc, Bauer:2010wj, Sahni:2014ooa, Gomez-Valent:2015pia, Akarsu:2019hmw, Akarsu:2019ygx, Ozulker:2022slu, Adil:2023exv}.
It is important to recall that $w(z)$ is not a physical observable, thus its divergence does not mean 
a physical pathology or an obvious constraint or failure of the model.
The origin of the pole is clear from the definition of the barotropic EoS, $w=p/\rho$, which 
occurs when the energy density of the quintom dark energy density turns to be zero, i.e., when the 
negative terms of $\rho_Q = \frac12(\dot{\phi}^2 -\dot{\psi}^2 +m_{\phi}^2\phi^2 +m_{\psi}^2\psi^2 
-2|\beta| \psi^2\phi^2)$ are relevant for certain values of $\beta$, such that $ \rho_Q= 0$, and hence 
$\rho_Q$ is able to change the sign to become negative. 
A negative energy density can be associated with the sign of the cosmological constant, the 
hypothesis of a negative mass, or just an effective energy density similar to the curvature case, 
see for instance \cite{Petit:2014ura, Visinelli:2019qqu, Calderon:2020hoc, Akarsu:2021fol}.
Regarding the Hubble function, it happens the opposite to the previous cases. 
If $H_{\Lambda{\rm CDM},0}$ is smaller than the Hubble parameter given by the quintom+$\beta$ 
model $H_{Q,0}$, that is $H_{\Lambda{\rm CDM},0}< H_{Q,0}$, then at some redshift it occurs 
that $H_{\Lambda{\rm CDM}}(z) > H_{Q}(z)$, and consequently the $D_H$ and $D_M$ traverse from 
bottom to top the $\Lambda$CDM line, as seen in the lower panels of the same figure. 

\section{Code and observations}\label{code an observations}

We perform the parameter estimation and provide observational constraints from the latest data on the 
free parameters of the quintessence, phantom, quintom, and quintom+$\beta$ dark energy models considering the 
potentials \eqref{eq:Vphi}, \eqref{eq:Vpsi}, \eqref{eq:Vtot} and \eqref{eq:quintom potential}, and discuss 
the model even further. 
In order to explore the parameter space, we use a modified version of dynesty, a library with several versions 
of the nested sampling algorithm. In conjunction, we utilize the SimpleMC cosmological parameter estimation 
code \cite{simplemc,aubourg2015}, which computes expansion rates and distances using the Friedmann equation 
to calculate the posterior distributions. Equipped with these tools, we can easily calculate the Bayesian 
evidence $\ln \mathcal{Z}$, an informative measure of the compatibility of the statistical model with the 
observed data, thereby allowing direct comparison of two cosmological models, $a$ and $b$, using the 
Bayes factor $B_{ab}\equiv \mathcal{Z}_a/\mathcal{Z}_b$, or equivalently the relative log-Bayes evidence 
$\ln B_{ab}\equiv \Delta \ln \mathcal{Z}$.
The model with smaller $|\ln \mathcal{Z}|$ is the preferred model, and to interpret the results, we refer to 
the Jeffreys' scale: a weak evidence is indicated by $0 \leq |\Delta\ln \mathcal{Z}|  < 1$, a moderate evidence 
$1 \leq | \Delta\ln \mathcal{Z}|  < 3$, a strong evidence by $3 \leq | \Delta\ln \mathcal{Z}|  < 5$, and a 
decisive evidence by $| \Delta\ln \mathcal{Z} | \geq 5$,  in favor of the model. 
For an extended review of cosmological parameter inference see \cite{padilla2019}.

To perform the parameter estimation, we consider data from cosmic chronometers (HD), Type Ia supernovae (SN), 
and Baryon Acoustic Oscillation measurements (BAO), which are detailed in the following list:

\begin{itemize}
	\item \textbf{HD}: Hubble distance measurements or cosmic chronometers are galaxies that evolve slowly 
    and allow direct measurements of the Hubble parameter $H(z)$. \changes{We use the most recent compilation that contains a covariance matrix from Reference \cite{moresco2020setting}}.
	\item \textbf{SN}: The SNeIa dataset used in this paper is the \changes{Pantheon+, a compilation of 1550 Type Ia supernovae within redshifts between $z=0.001$ and $z=2.26$ \cite{brout2022pantheon+}}.
	\item \textbf{BAO}: High-precision Baryon Acoustic Oscillation measurements (BAO) at different redshifts up 
    to $z<2.36$. We make use of the BAO data from SDSS DR12 Galaxy Consensus, BOSS 
    DR14 quasars (eBOSS), Ly-$\alpha$ \changes{DR16} cross-correlation, 
    Ly-$\alpha$ \changes{DR16} auto-correlation, Six-Degree Field Galaxy Survey (6dFGS) and  SDSS Main Galaxy Sample (MGS) \cite{eBOSS:2020yzd}. 
    %
\end{itemize}

Throughout the analysis, we assume a flat FLRW universe, and flat priors over our sampling parameters: 
 $\Omega_{{\rm m}0}=[0.05,1.0]$ for the pressureless matter density parameter today, 
$\Omega_{{\rm b}0} h^2=[0.02,0.025]$ for the physical baryon density parameter and $h=[0.4,1.0]$ for the reduced 
Hubble constant; additional to these parameters, for the quintom model we have $m_{\phi} = [0.0, 4.0]$ the 
quintessence field mass, $m_{\psi}=[0.0,3.0]$ the phantom field mass and $\beta =[-10.0, 10.0]$ for the coupling 
parameter, but when using the combination of all the datasets (BAO+HD+SN) we have used $\beta =[-2.0, 4]$.

\section{Results}\label{results}
\begin{figure}[t!]
\captionsetup{justification=raggedright,singlelinecheck=false,font=small}
\begin{center}
\includegraphics[trim = 1mm  0mm 1mm 1mm, clip, width=8.cm, height=7cm]{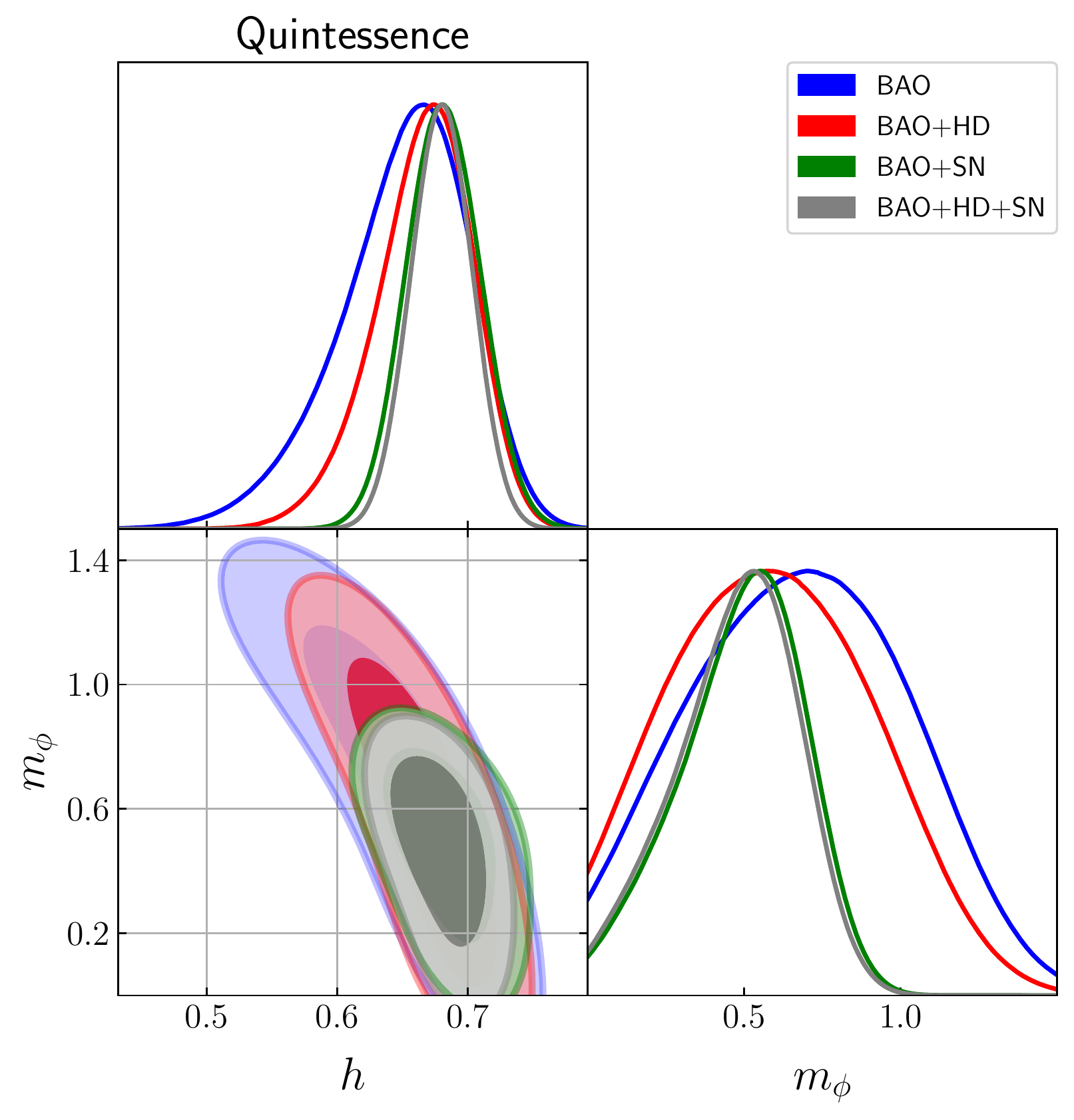} 
\includegraphics[trim = 1mm  0mm 1mm 1mm, clip, width=8.cm, height=7cm]{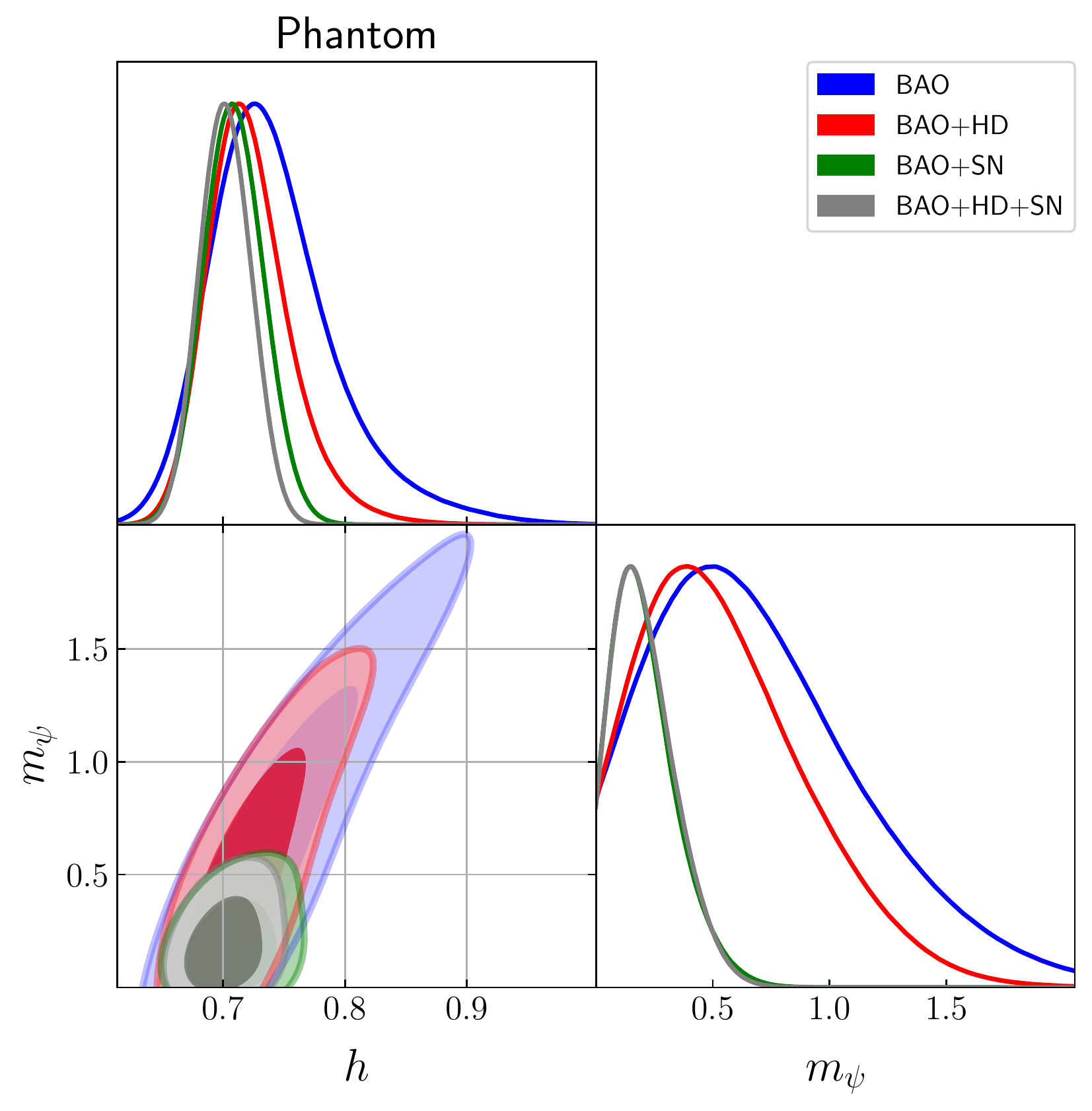} 
\end{center}
\caption[Quintess]{One- and two-dimensional (68\% and 95\% CLs) marginalized posterior distributions 
for the free parameters of quintessence (left) and phantom (right) dark energy models using BAO (blue), 
BAO+HD (red), BAO+SN (green) and BAO+HD+SN (gray). 
}
\label{Fig:density_quintessphantom2}
\end{figure}

\begin{figure}[t!]
\captionsetup{justification=raggedright,singlelinecheck=false,font=small}
\begin{center}
\includegraphics[trim = 1mm  0mm 1mm 1mm, clip, width=8.2cm, height=8cm]{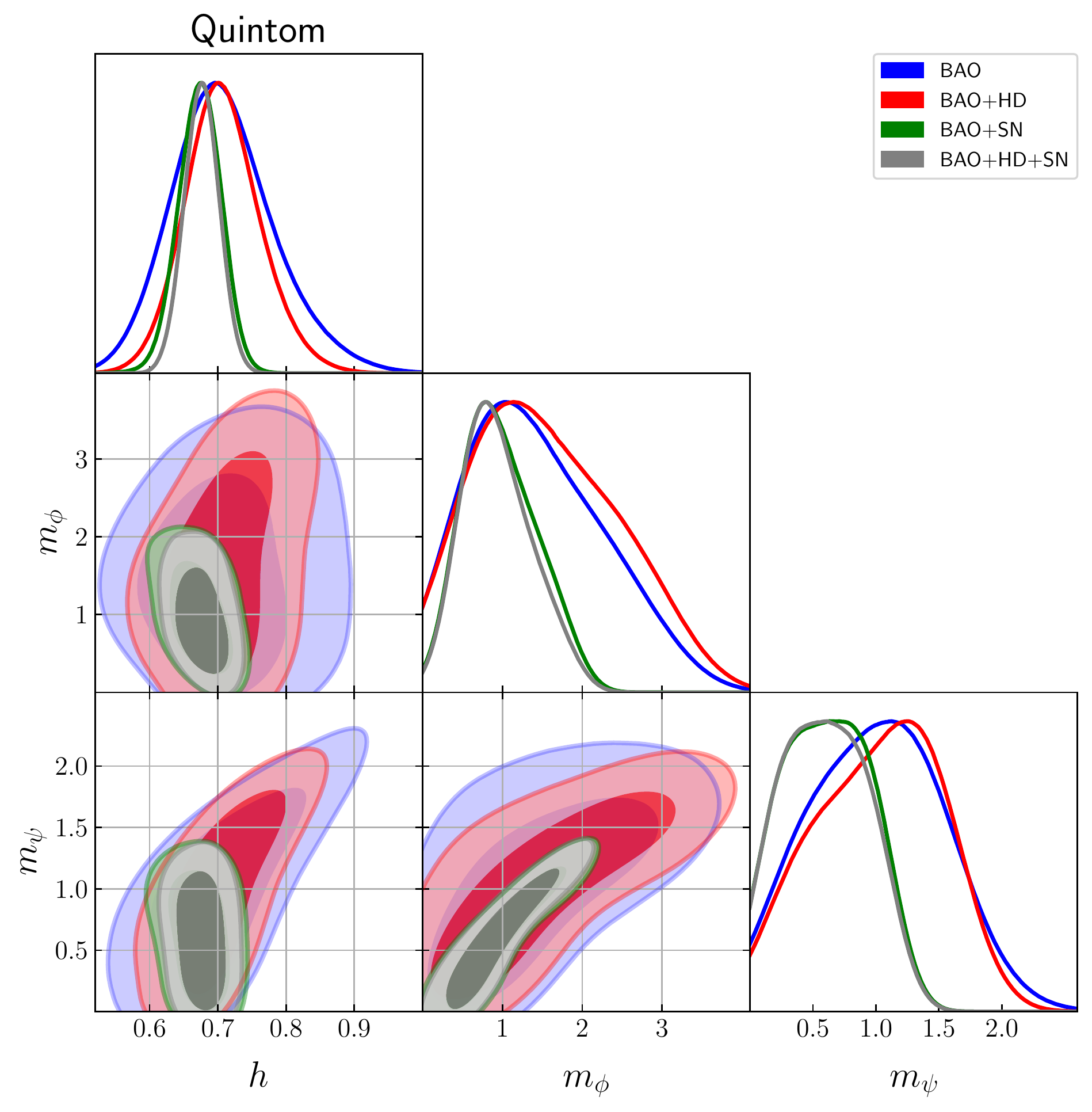}
\includegraphics[trim = 1mm  0mm 1mm 1mm, clip, width=9.3cm, height=8.5cm]{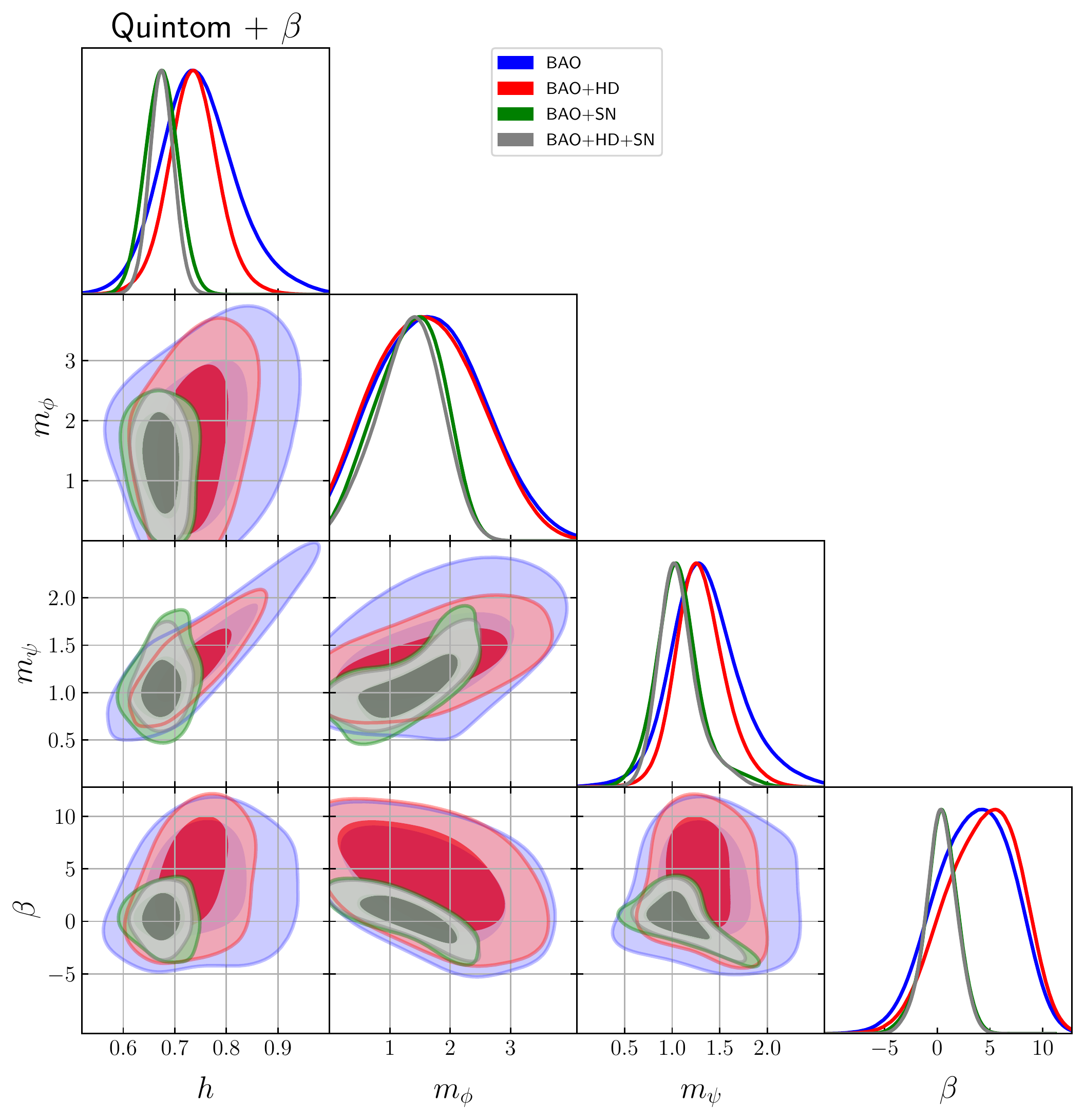}
\end{center}
\caption[Quintom]{One- and two-dimensional (68\% and 95\% CLs) marginalized posterior distributions for 
the free parameters of quintom (left) and quintom$+\beta$ (right) dark energy models using BAO (blue), 
BAO+HD (red), BAO+SN (green) and BAO+HD+SN (gray).
}
\label{Fig:density_quintom2}
\end{figure}

The main results of our analysis are shown in Table \ref{tab:nestedestimation} where we report the 
constraints of the model parameters $\Omega_m$, $h$, $m_{\phi}$, $m_{\psi}$, $\beta$ along with 68\% 
confidence level (CL), for all different combinations of the data sets: HD, SN, BAO, HD+SN, BAO+HD, 
BAO+SN, BAO+HD+SN. We also include the results of $\Lambda$CDM, as the reference model.
Additionally, the same Table displays the best-fit, $-2\ln\Like_{\rm max}$, along with the 
log-Bayesian evidence, $\ln \mathcal{Z}$,  
from the nested sampling algorithm, with the number of live-points selected using the general rule 
$50 \times ndim$~\cite{dynesty}, where $ndim$ is the number of parameters to be sampled from. 
Complementary to the table, in Figures \ref{Fig:density_quintessphantom2} and \ref{Fig:density_quintom2}, 
we display the two-dimensional marginalized distributions (the inner and external contours are for 
68\% and 95\% CLs, respectively) and the constraints in the form of one-dimensional marginalized posterior 
distributions: figure \ref{Fig:density_quintessphantom2} corresponds to quintessence and phantom dark energy, 
and figure \ref{Fig:density_quintom2} for quintom and quintom$+\beta$, in both cases we include the 
combination of datasets that provided the most constraining power.
\\

In general, and among datasets, the Quintessence and Phantom models are statistically consistent with 
$\Lambda$CDM, however, there are some important points to mention. The quintessence model presents a nearly 
negligible improvement to describe datasets, that is, the comparison relative to the reference 
model \footnote{Throughout the 
results we use $\Lambda$CDM as the reference model to compare with model $i$,
thus $-2\ln\Delta \Like_{\rm max}\equiv -2\ln (\Like_{\rm \Lambda CDM,max}/\Like_{i,{\rm max}}$).}
is small $-2\ln \Delta \Like_{\rm max} \lesssim 1$, except {when both BAO and SN 
appear in the combined dataset, for these combinations, the improvement of the fit yield 
$-2\ln\Delta \Like_{\rm max}\leq 2.85$.
This can be understood by the lower $m_{\phi}$ values, which are accompanied by distinguishable values 
of $h$ compared to their counterpart in the $\Lambda$CDM model. Additionally from left panel 
of Figure~\ref{Fig:density_quintessphantom2} is evident that these datasets impose
the tightest parameter constraints.
Notice that the HD+SN combination gives the lowest $m_{\phi}$ value, nevertheless, for this reason, 
it is more capable of mimicking
$\Lambda$CDM, carrying a barely different change in $h$ and a very similar fit. Consequently, 
the advantage of the additional parameter is lost.  
From the calculation of the Bayesian evidence, we can interpret that while $\Lambda$CDM is 
preferred over the quintessence model, this preference is generally weak for most of the data sets. 
However for SN and HD+SN, the evidence in favor of $\Lambda$CDM is definite but not strong.}

{In contrast to the quintessence model, the phantom model exhibits a positive 
correlation between $h$ and $m_{\psi}$ when considering the BAO and extra data, 
see left panel of Figure~\ref{Fig:density_quintessphantom2}. However, in this model, the value of 
$-2\ln \Delta \mathcal{L}_{\rm max}$ does not show improvement compared to $\Lambda$CDM,
even when for BAO+SN and BAO+HD+SN the constraints on the scalar field mass are more restrictive 
than the found for quintessence.
Analyzing the Bayesian evidence, it suggests that while the phantom model, can be weakly or 
definitely less favored than $\Lambda$CDM, there is an exception for the HD sample, where it 
is indeed weakly favored. Notably, in this instance the parameter adjustment for $m_{\psi}$ 
exhibits the biggest value for this model.}
\\

{On the other hand, when the two fields are incorporated into the quintom model, notice that with the inclusion of the BAO data, there is a positive correlation among the masses of the fields, 
but the correlation of the individual masses with the hubble parameter is essentially lost (see left panel of 
Figure \ref{Fig:density_quintom2}). }
{For all data combinations, the parameter adjustments favor $m_{\phi} > m_{\psi}$
indicating that the quintom model prefers to be dominated by the quintessence branch.
Similar to the quintessence model, the fit is improved when the BAO and SN data appear in the dataset; 
in these cases $-2\ln\Delta \Like_{\rm max}\leq 3.58$, 
and their corresponding $h$ value is lower than that obtained for $\Lambda$CDM.
However, the penalization carried by the extra parameters is more evident when analyzing the Bayes factors.
According to this criterion, it is notable that when only HD data is employed, both models have the same 
Bayes factor and then both are equally favored.
For the BAO and BAO+HD combination, the quintom model is weakly disfavored compared to the standard model. 
Furthermore, for the remaining data combinations, the preference over $\Lambda$CDM is enhanced and now definitive.
It is worth nothing that under the Bayesian evidence, even the BAO+SN and BAO+HD+SN combinations, 
which previously seemed to provide better fits, now are disfavored. 
}
\\

\begin{figure}
\captionsetup{justification=raggedright,singlelinecheck=false,font=small}
 \makebox[11cm][c]{
            \includegraphics[trim=0mm 0mm 0mm 0mm, clip, width=6cm, height=6cm]{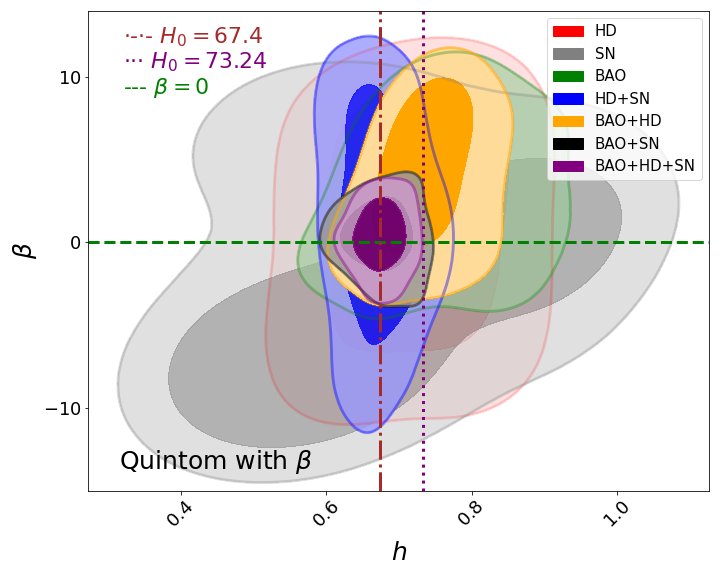}
            \includegraphics[trim=0mm 0mm 0mm 0mm, clip, width=6.5cm, height=6cm]{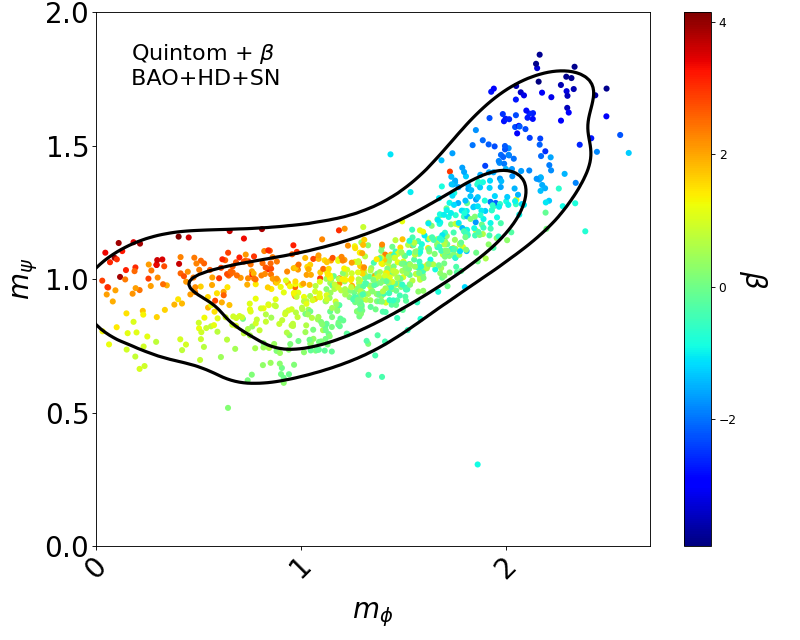}
            }
    \caption{Quintom$+\beta$ model. \textit{Left panel}: two-dimensional marginalized posterior distributions 
    (68\% and 95\% CLs) $\beta-h$ plane for different combinations of datasets;  the dashed lines correspond to 
    the $\Lambda$CDM model: $H_0 = 73.24 ;$ km s$^{-1}$ Mpc$^{-1}$, coming from the Cepheid variables \cite{riess20162}; 
    and $H_0 = 67.40 ;$ km s$^{-1}$ Mpc$^{-1}$, measured by the Planck mission \cite{aghanim2020planck}. 
    \textit{Right panel}: scatter plot in the $m_{\phi}-m_{\psi}$ plane for different values of $\beta$ (color bar), 
    for BAO+HD+SN. 
    }
    \label{fig:quintom_beta}
\end{figure}

Now let us focus on the novel model, quintom$+\beta$. The first point to highlight is that the $\beta$ 
parameter is bounded by the different datasets, and even though the BAO and its 
combinations indeed provide tight constraints, with a slight preference for positive values of 
$\beta$, {except for the SN data}, all
datasets and its combinations point out that $\beta=0$ is  statistically admissible
(the two-dimensional marginalized distributions of $\beta$ and $h$ are shown in the right panel of 
figure \ref{fig:quintom_beta} for different combinations of datasets).
Despite this fact, 
some important features come up with the introduction of this parameter. {Some 
points to remark concerning the quintom+$\beta$ model are that, 
once again, the mass of the quintessence field is larger than the mass of the phantom field. 
This implies that, in this model as well, a preference exists for quintessence domination. 
It is also important to note that the masses of the fields are generally larger than those in the quintom model.
In fact, for quintessence, all of them are greater than 1.}
In the right panel of Figure \ref{Fig:density_quintom2} we show the constraints of the 
quintom$+ \beta$ model using BAO and additional datasets.
A similar correlation occurs with the mass of the phantom and quintessence field
(compared to the standard quintom model without coupling). This is an important point to stress out because 
now the datasets are in favor of $m_\psi$ different from zero with more than $2\sigma$ CL, in fact, the tightest constraint yields, 
$m_\psi=1.067 \pm 0.214$ (BAO+HD+SN),  which also yields to 
higher values of the quintessence mass $m_\phi= 1.305 \pm 0.533$ (BAO+HD+SN). To have a closer look of the 
correlations among the quintom$+\beta$ parameters, middle 
panel of figure \ref{fig:quintom_beta} shows the 2D marginalized posterior distributions of $m_\phi, m_\psi$
color coded with $\beta$ values.
{In this case, the most significant improvement to the fit occurs again when we consider the BAO+SN dataset (with $-2\ln\Delta \Like_{\rm max}= 3.56$ ). However, the results for the BAO+SN+HD combination are very close ($-2\ln\Delta \Like_{\rm max}= 3.28$ ).}
The inclusion of this coupling produces a difference in favor of the model  that contributes to diminishing the BAO tension created between 
low redshift (galaxies) and high (Ly-$\alpha$) data, explored in \cite{aubourg2015, Akarsu:2019hmw}. 
Even though this model contains three extra parameters, the Bayes factor, with respect to $\Lambda$CDM, {highlights this aspect by imposing a more significant penalty compared to all the other models under consideration.
However, this penalty is not high enough to discard the quintom+$\beta$ model, as
$|\Delta \ln \mathcal{Z}|\leq 3.84$.
Specifically the values of $|\Delta \ln \mathcal{Z}|$ indicate that when considering only HD data it is weakly disfavored; it is definitively disfavored for SN, BAO and
BAO+SN combinations; and strongly disfavored by HD+SN, BAO+SN, and BAO+HD+SN dataset combinations. In a work in progress, for the quintom model with coupling, we are testing different potentials along with the 
Planck dataset (including linear perturbations), however for now, we are interested in the background cosmology.}
\\
    
\begin{figure}[t!]
\captionsetup{justification=raggedright,singlelinecheck=false,font=small}
 \makebox[11cm][c]{
            \includegraphics[trim=0mm 0mm 10mm 0mm, clip, width=9cm, height=5.5cm]{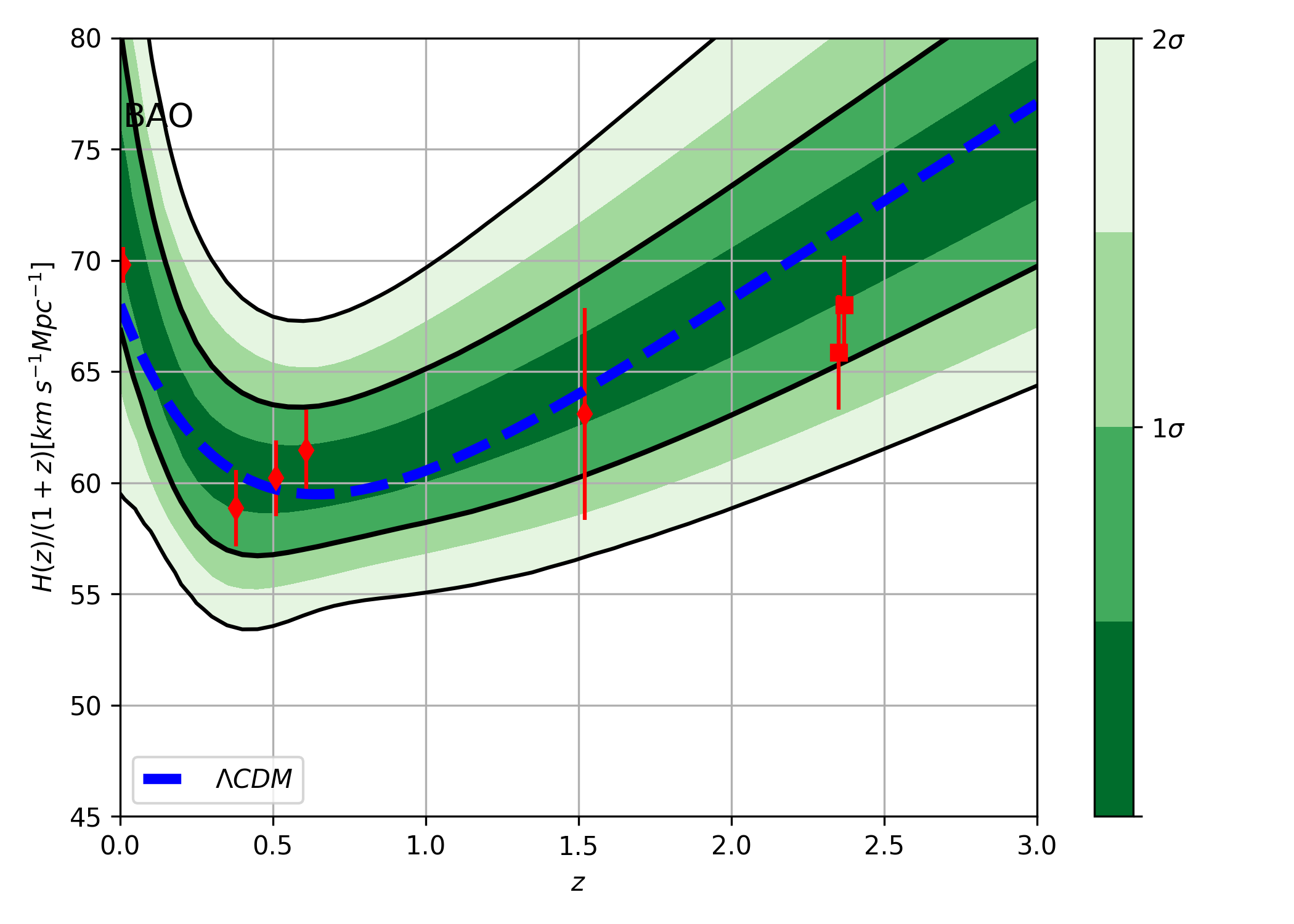}
            \includegraphics[trim=0mm 0mm 10mm 0mm, clip, width=9cm, height=5.5cm]{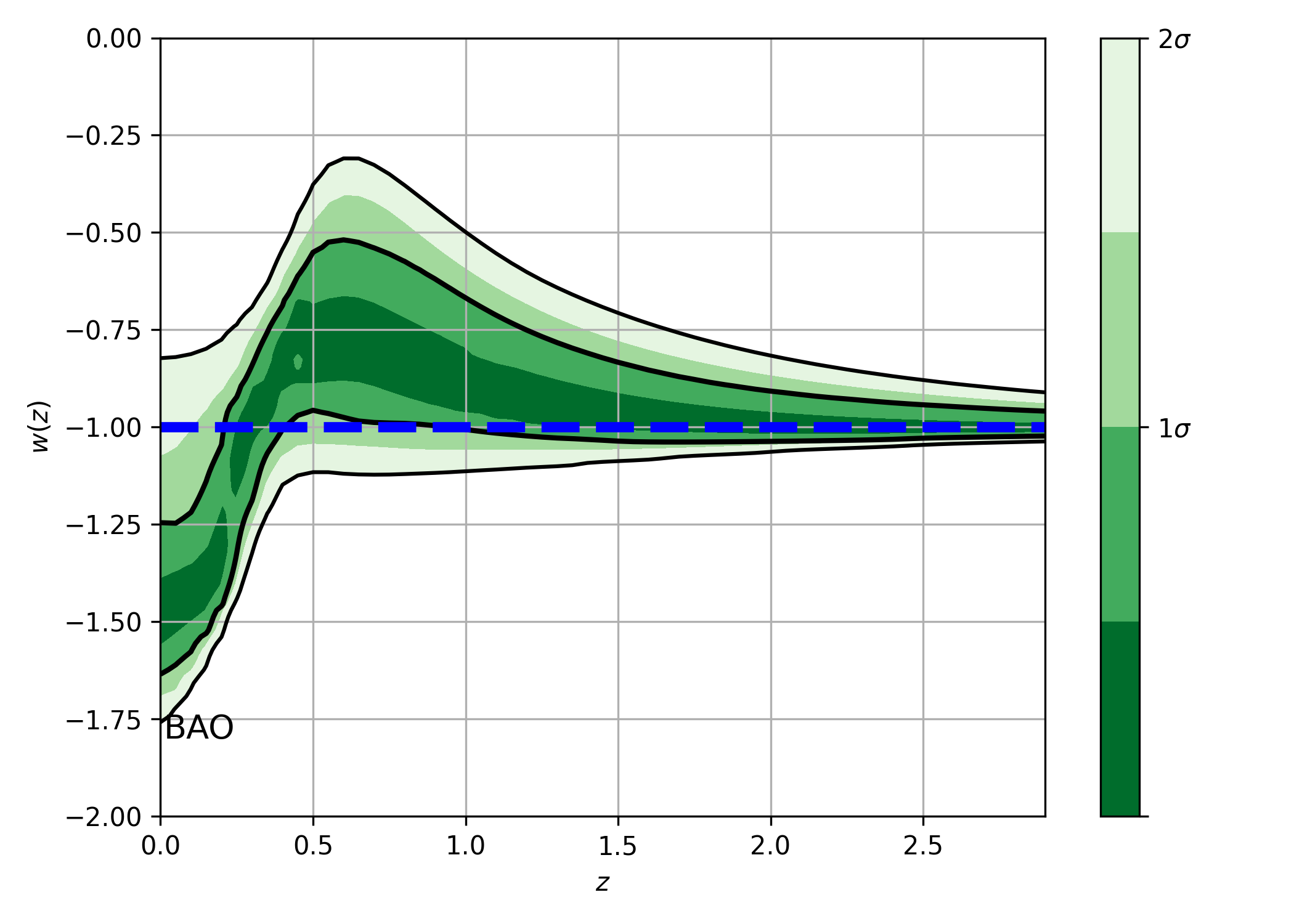}}
 \makebox[11cm][c]{
            \includegraphics[trim=0mm 0mm 10mm 0mm, clip, width=9cm, height=5.5cm]{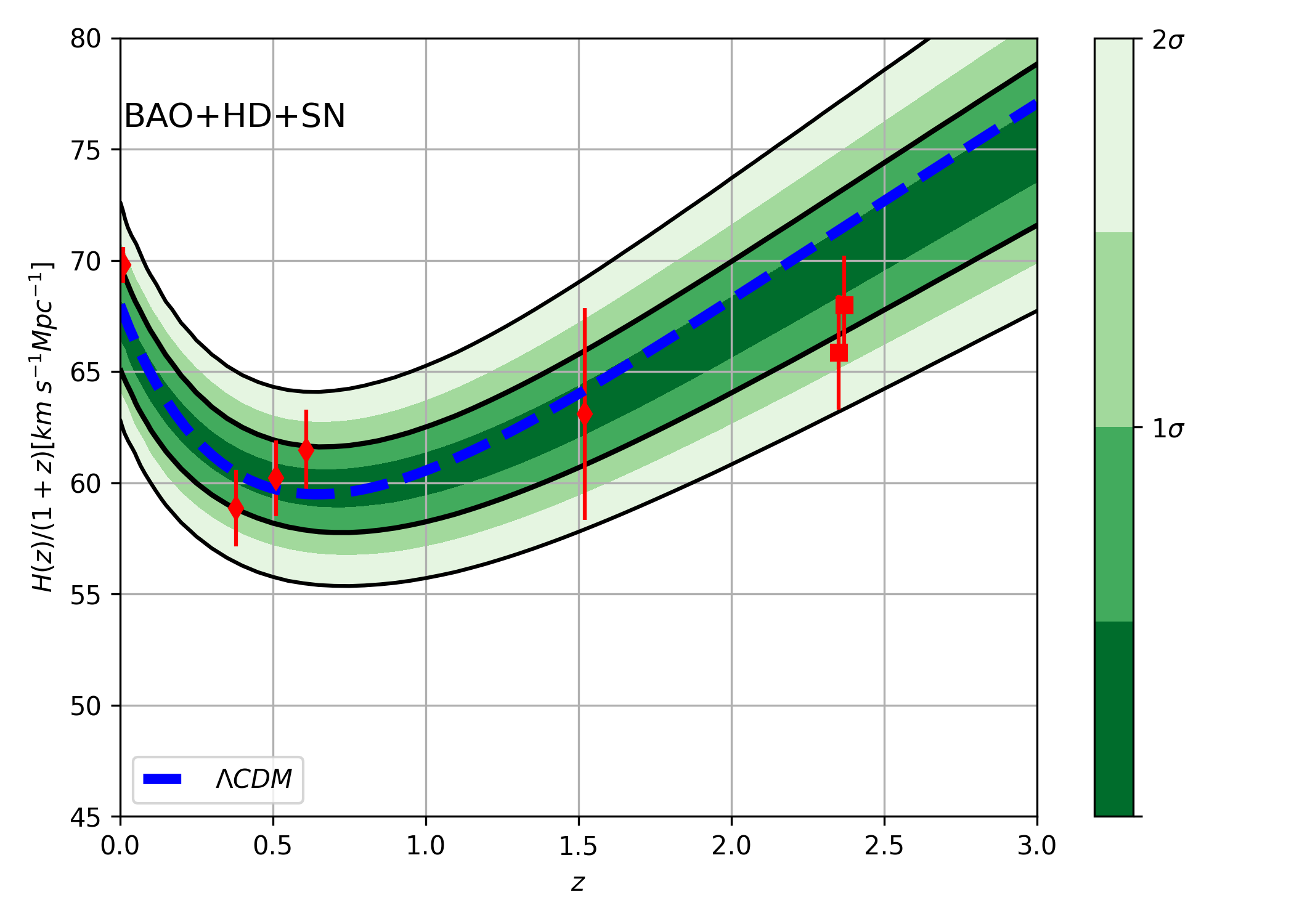}
            \includegraphics[trim=0mm 0mm 10mm 0mm, clip, width=9cm, height=5.5cm]{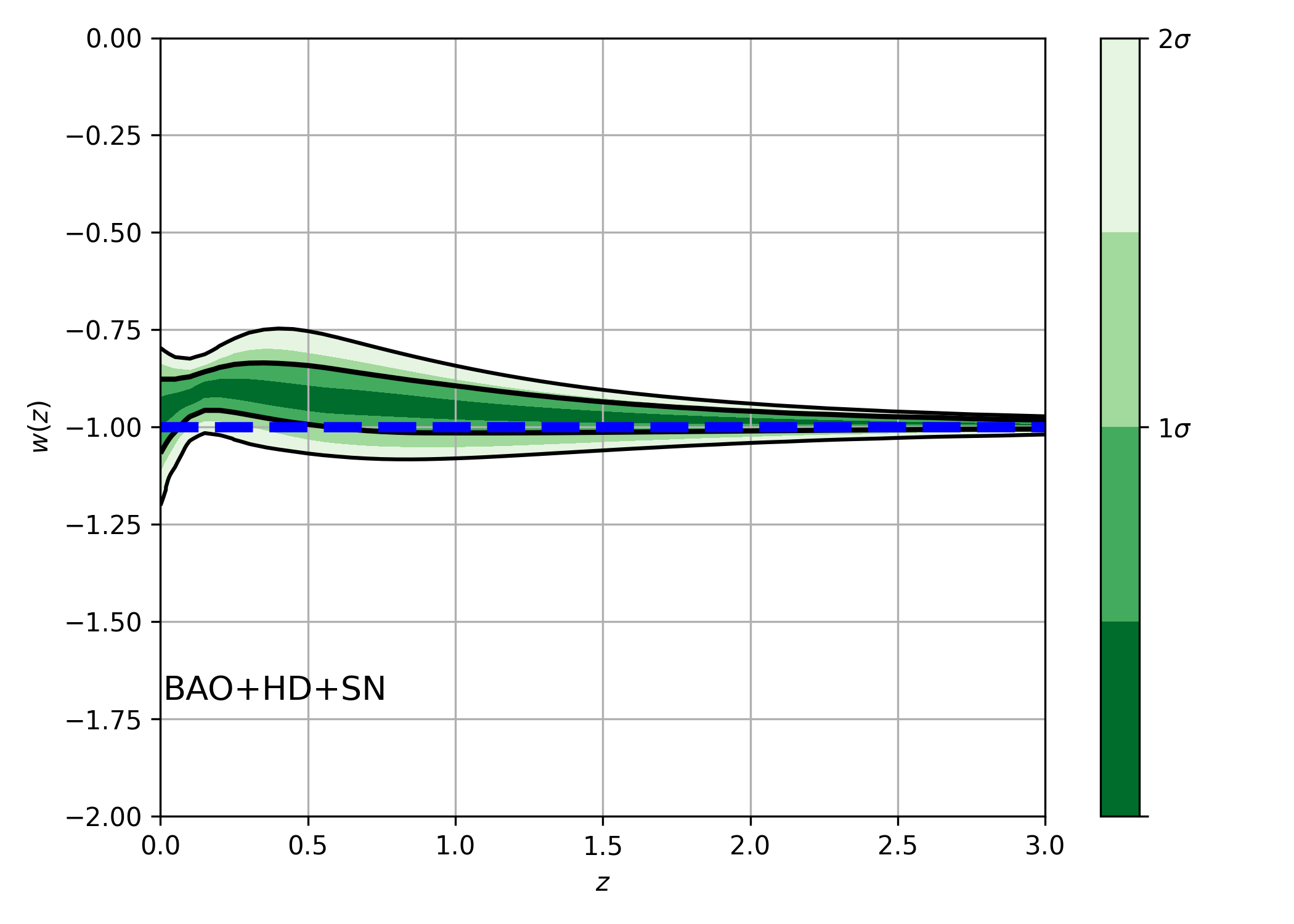}
            }

    \caption{Cosmological functions derived from the posterior distributions using the quintom$+\beta$ model. 
    \textit{Left}: $H(z)/(1 + z)$ and \textit{right} the EoS parameter $w(z)$, using BAO (\textit{top}) and  
    BAO+HD+SN (\textit{bottom}) datasets. The blue dashed lines represent the value of $\Lambda$CDM with the same 
    parameter estimation as shown in the Table \ref{tab:nestedestimation}. The red error bars correspond to 
    $H_0 = 69.8 \pm 0.8$ km s$^{-1}$ Mpc$^{-1}$ from the TRGB  \cite{freedman2019carnegie}, BAO Galaxy consensus 
    ($z \sim 0.5$) \cite{alam2017clustering}, BOSS DR14 quasars (eBOSS, $z=1.52$) \cite{ata2018clustering} and 
    Ly-$\alpha$ \changes{DR16 and cross-correlation ($z=2.35$) and auto-correlation ($z=2.37$) \cite{eBOSS:2020yzd}.}}
    \label{fig:quintom_couple}
\end{figure}

Once we have performed the parameter estimation, we are able to plot some derived probability distribution
functions in order to look where are located the main deviations from the $\Lambda$CDM model. For instance,  
the left panel of figure \ref{fig:quintom_couple} displays the Hubble function $H(z)/(1 + z)$ and the right panel
the dark energy equation of state (equation \eqref{eq:eos}). In this figure we present the results using BAO (top)
and the combined BAO+HD+SN dataset (bottom), and also include the TRGB $H_0$ and BAO data points for 
comparison (red error bars). 
The solid lines represent 1$\sigma$ and 2$\sigma$ CLs and darker color means more likely as shown in the color bar; 
for comparison, the dashed blue line corresponds to the $\Lambda$CDM prediction. We observe there is a  shift in 
the amplitude to lower values of $H(z)$, being clear for the case of BAO where almost the entire 1$\sigma$ region 
is lower than $\Lambda$CDM, and for the case of combined data this shift becomes small, as the constraints 
increased. Regarding the effective DE EoS, its best-fit value at $z=0$ is located below $w=-1$ 
(allowing small deviations incline to the phantom region), then it increases to cross the PDL and reaching a 
maximum value at about $z=0.5$, where the BAO-galaxy points are located, (also observed as an inverted bump 
in the $H(z)$ figure). Notice that at this redshift the quintom$+\beta$ model deviated by more than the 
$2\sigma$ region from the cosmological constant. After $w(z)$ has achieved the maximum value, it decreases 
to slightly cross back the PDL again; this may be an intent to fit the BAO-Ly$\alpha$ as well.  
For the case of joint data analysis we have a similar behavior but more constrained, with $w(z=0) \approx -1.1$ 
with a more restricted range but also favors the phantom region, the maximum is smaller too and throughout 
the redshift range $w=-1 $ is acceptable.
We must emphasize that the behavior presented by the EoS of the quintom$+\beta$ model with the values obtained 
from observational constraints (right panel of \ref{fig:quintom_couple}) is qualitatively similar to that obtained 
from model-independent reconstructions \cite{Vazquez:2012cen, Hee:2016nho, Zhao:2017cud, Wang:2018fng}, 
showing that the quintom with coupling is a plausible option to model the most recent observational 
results of dark energy.

\section{Conclusions and Discussion}\label{conclusions}

In this work, we have analyzed models of dark energy as minimally coupled scalar fields, specifically a quintessence 
and phantom models with quadratic potential eqs. \eqref{eq:Vphi} and \eqref{eq:Vpsi}, a quintom model as the 
sum of the quadratic potentials of the previous two fields Eq. \eqref{eq:Vtot}; and a novel proposal, dubbed 
quintom$+\beta$ with sum of the quadratic potentials and an interaction term Eq. \eqref{eq:quintom potential} 
inspired by \cite{PerezLorenzana:2007qv}. 
Although quintessence and phantom with quadratic potential, as well as quintom as the sum of these, have been 
well-known models for some time, in order to make an update, in this work we constrained the parameters using 
late time data. Our main focus is to study the new proposal quintom$+\beta$, which enriches the quintom model 
in a simple way by adding only one  extra degree of freedom to the usual non-minimal quintom model (the simplest 
coupling between the quintessence and phantom fields). 
We show that one of the main features of quintom$+\beta$ is that it produces a wavelike EoS (see figure 
\ref{fig:mquin-mphan}) similar to the one reported in other works in other contexts \cite{Tamayo:2019gqj}. 
When constraining the parameters, quintom$+\beta$ fits very well to the background cosmological data, specially 
for the {SN,  which exhibits a preferred negative $\beta$ value} and results in an
improvement of $-2\ln\Delta \Like_{\rm max}= 5.19$, compared to $\Lambda$CDM.
This result leads to an EoS that starts, at early times, slightly below $-1$, then increases smoothly peaking at 
$z\approx 0.5$ and then decreases to end with a phantom value at present time $ w(z=0)\approx-1.2$. 
Using the values of best fit, the CL for $H(z)$ and the EoS were calculated, the best fit of the parameters provides 
an EoS qualitatively similar to that reported by non-parametric reconstructions of dark energy \cite{Zhao:2017cud} 
(see figure top right \ref{fig:quintom_couple}). As has already been said, this is precisely the behavior obtained 
from reconstructions, which allows us to say that quintom$+\beta$ with only one additional parameter to the two 
necessarily associated with the quintessence field and phantom, may reproduce the possible nature of the EoS 
obtained from observations. This makes the proposal of this work interesting, which motivates a more careful 
exploration of quintom models with potentials that couple the quintessence and phantom fields. It seems that 
interactions between fields through the product of powers of the fields, $V_{\rm int}\propto \phi^n \psi^m$, may 
play an important role in describing the dynamics of dark energy and may be a more economical way to add dynamics 
(several crosses to the PDL) instead of potentials with elaborated functions with a vaguely clear justification.

For four scalar field dark energy models (quintessence, phantom, quintom, quintom$+\beta$) and for 
$\Lambda$CDM (for comparison), the cosmological and model parameters were constrained using HD, SN and BAO 
data, and its combinations; in addition to calculating the Bayesian evidence. 
The observational constraints of the quintessence and phantom potential parameters associated to the mass 
of the fields, $m_\phi$ and $m_\psi$, give values for both around $m_{\phi, \psi} \approx 0.2-1.5$ and 
with a slight correlation with $h$ ($m_\phi$ directly proportional and $m_\psi$ inversely proportional). 
For the quintom model also the parameters are of the same order and slightly bigger than quintessence and 
phantom, $m_{\phi, \psi} \approx 0.6-1.8$; but in this case without correlation with $h$. For the proposed 
model quintom$+\beta$, the mass parameters are both $m_{\phi, \psi} \approx 1.0-2.06$ and the coupling 
parameter $\beta\approx -3.1-3.6$; it is noteworthy that in all cases the error bars of the constraints are of 
the order of the size for the mass parameters and even larger for the $\beta$ parameter. 
A natural improvement is to consider the CMB data and for a more detailed study consider linear perturbations. 

The proposal to model the dark energy with quintessence and phantom scalar fields (to allow PLD crossing) with a 
potential with interacting term (to allow several crossings) seems to be reasonable since it is a flexible model 
that allows a dynamical EoS requiring three parameters (one for each field and one for the interaction) and 
with a simple functional form for the potential (polynomials). The quintom models are still valid as a candidate 
for dark energy; a deeper and more careful study on the origin of scalar fields is needed as well as better observational constraints.

\begin{table}[t!]
\captionsetup{justification=raggedright,singlelinecheck=false,font=small}
\scriptsize
\begin{tabular}{p{2.0cm}p{2.2cm}p{2.2cm}p{2.2cm}p{2.2cm}p{2.2cm}p{2.2cm}}
\cline{1-7}
\noalign{\smallskip}
Parameter & Datasets \quad\quad & Quintessence \quad\quad &  Phantom\quad\quad  &  Quintom \quad\quad & 
Quintom$+\beta$ \quad\quad  & $\Lambda$CDM \quad\quad \\
\hline
\\
 $\Omega_m$ & HD & $0.362 \pm 0.120$ & $0.346 \pm 0.100$ & $0.332 \pm 0.103$ & $0.346 \pm 0.117$ & $0.368 \pm 0.108$ \\
 \vspace{0.15cm}
& SN & $0.306 \pm 0.032$ & $0.347 \pm 0.026$ & $0.309 \pm 0.044$ & $0.354 \pm 0.083$ &  $0.332 \pm 0.018$ \\
 \vspace{0.15cm}
 & BAO &  $0.300 \pm 0.021$ & $0.285 \pm 0.020$ & $0.276 \pm 0.030$ & $0.256 \pm 0.033$ & $0.292 \pm 0.019$ \\
 \vspace{0.15cm}
 & HD+SN &  $0.311 \pm 0.029$ & $0.346 \pm 0.024$ & $0.311 \pm 0.045$ & $0.314 \pm 0.084$ & $0.332 \pm 0.018$\\
 \vspace{0.15cm}
 & BAO+HD &  $0.300 \pm 0.019$ & $0.287 \pm 0.018$ & $0.276 \pm 0.030$ & $0.252 \pm 0.029$ & $0.292 \pm 0.018$ \\
 \vspace{0.15cm}
 & BAO+SN &  $0.300 \pm 0.016$ & $0.315 \pm 0.014$ & $0.296 \pm 0.018$ & $0.295 \pm 0.018$ & $0.313 \pm 0.014$ \\ 
 \vspace{0.15cm}
& BAO+HD+SN &  $0.301 \pm 0.015$ & $0.314 \pm 0.014$ & $0.298 \pm 0.016$ & $0.297 \pm 0.016$ & $0.312 \pm 0.013$ \\
\hline
\\
$h$  & HD & $0.630 \pm 0.062$ & $0.723 \pm 0.083$ & $0.722 \pm 0.089$ & $0.723 \pm 0.088$ & $0.661 \pm 0.057$ \\
 \vspace{0.15cm}
 & SN &  $0.755 \pm 0.162$ & $0.737 \pm 0.166$ & $0.800 \pm 0.141$ & $0.671 \pm 0.173$  & $0.699 \pm 0.174$\\
 \vspace{0.15cm}
 & BAO &  $0.652 \pm 0.048$ & $0.739 \pm 0.050$ & $0.707 \pm 0.074$ & $0.747 \pm 0.073$  & $0.699 \pm 0.022$   \\
 \vspace{0.15cm}
 & HD+SN &  $0.677 \pm 0.040$ & $0.674 \pm 0.038$ & $0.677 \pm 0.038$& $0.673 \pm 0.038$ & $0.674 \pm 0.040$  \\
 \vspace{0.15cm}
 & BAO+HD &  $0.667 \pm 0.036$ & $0.721 \pm 0.033$ & $0.707 \pm 0.055$ & $0.736 \pm 0.048$ & $0.696 \pm 0.019$\\
 \vspace{0.15cm}
 & BAO+SN &  $0.681 \pm 0.027$ & $0.709 \pm 0.024$ & $0.674 \pm 0.031$ & $0.674 \pm 0.031$ & $0.703 \pm 0.023$ \\
 \vspace{0.15cm}
& BAO+HD+SN &  $0.680 \pm 0.023$ & $0.702 \pm 0.020$ & $0.676 \pm 0.025$ & $0.675 \pm 0.023$ & $0.697 \pm 0.019$ \\
\hline
\\ 
$m_\phi$& HD &  $0.841 \pm 0.524$
 & $- $ & $1.780 \pm 1.152$ & $2.063 \pm 1.062$ &  \\
 \vspace{0.15cm}
 & SN &  $0.493 \pm 0.311$ & $-$ & $1.000 \pm 0.660$ & $1.383 \pm 0.675$ & \\
 \vspace{0.15cm}
 & BAO &  $0.664 \pm 0.346$ & $-$ & $1.386 \pm 0.867$ & $1.636 \pm 0.880$ & \\
 \vspace{0.15cm}
 & HD+SN & $0.409 \pm 0.241$ & $-$ & $0.852 \pm 0.534$ & $1.461 \pm 0.814$ &  \\
 \vspace{0.15cm}
 & BAO+HD &  $0.573 \pm 0.326$ & $-$ & $1.514 \pm 0.926$ & $1.579 \pm 0.864$ &  \\
 \vspace{0.15cm}
 & BAO+SN &  $0.492 \pm 0.195$ & $-$ & $0.971 \pm 0.466$ & $1.313 \pm 0.553$ & \\
 \vspace{0.15cm}
& BAO+HD+SN & $0.473 \pm 0.194$ & $-$ & $0.933 \pm 0.443$ & $1.305 \pm 0.533$ & \\

\hline
\\
 $m_\psi$& HD &  $-$ & $1.496 \pm 0.846$ & $1.662 \pm 0.786$ & $1.788 \pm 0.751$  & \\
 \vspace{0.15cm}
 & SN &  $-$ & $0.407 \pm 0.307$ & $0.713 \pm 0.448$ & $1.004 \pm 0.515$ & \\
 \vspace{0.15cm} 
 & BAO &  $-$ & $0.635 \pm 0.466$ & $0.989 \pm 0.534$ & $1.354 \pm 0.370$ & \\
 \vspace{0.15cm}
 & HD+SN &  $-$ & $0.360 \pm 0.243$ & $0.628 \pm 0.355$ & $0.971 \pm 0.360$ &  \\
 \vspace{0.15cm}
 & BAO+HD &  $-$ & $0.502 \pm 0.364$ & $1.014 \pm 0.506$ & $1.297 \pm 0.260$ &  \\
 \vspace{0.15cm}
 & BAO+SN &  $-$ & $0.189 \pm 0.140$ & $0.601 \pm 0.354$ & $1.068 \pm 0.247$ & \\
 \vspace{0.15cm}
& BAO+HD+SN &  $-$ & $0.191 \pm 0.137$ & $0.586 \pm 0.350$ & $1.067 \pm 0.214$ &  \\

\hline
\\
$\beta$ & HD &  $-$ & $-$ & $-$ & $1.244 \pm 5.413$ & \\
 \vspace{0.15cm}
& SN &  $-$ & $-$ & $-$ & $-3.120 \pm 5.525$ & \\
 \vspace{0.15cm}
 & BAO &  $-$ & $-$ & $-$ & $3.590 \pm 3.504$ &   \\
 \vspace{0.15cm}
 & HD+SN &  $-$ & $-$ & $-$ & $1.041 \pm 4.771$ & \\
 \vspace{0.15cm}
 & BAO+HD &  $-$ & $-$ & $-$ & $4.344 \pm 3.395$ & \\
 \vspace{0.15cm}
 & BAO+SN &  $-$ & $-$ & $-$ & $0.382 \pm 1.548$ & \\
 \vspace{0.15cm}
& BAO+HD+SN &  $-$ & $-$ & $-$ & $0.375 \pm 1.476$ & \\

\hline
\\
$-2\ln\Like_{\rm max}$  & HD &  $6.11$ & $5.40$ & $4.87$ & $4.39$ & $6.11$   \\
 \vspace{0.15cm}
& SN &  $1402.88$ & $1403.12$ & $1402.57$ & $1397.92$ & $1403.11$  \\
 \vspace{0.15cm}
& BAO &  $9.06$ & $ 9.45$ & $7.80$ & $7.34$ & $9.46$  \\
 \vspace{0.15cm}
 & HD+SN &  $1409.13$ & $1409.25$ & $1409.13$ & $1407.95$ & $1409.23$  \\
 \vspace{0.15cm}
 & BAO+HD &  $15.63$ & $15.87$ & $14.15$ & $13.41$ & $15.87$ \\
 \vspace{0.15cm}
 & BAO+SN &  $1412.14$ & $1414.97$ & $1411.41$ & $1411.43$ & $1414.99$ \\
 \vspace{0.15cm}
& BAO+HD+SN &  $1418.59$ & $1421.4$ & $1418.09$ & $1418.11$ & $1421.39$
\\
\hline
\\
$\ln \mathcal{Z}$& HD &  $-8.75 \pm 0.18$ & $-7.67 \pm 0.16$ & $-7.76 \pm 0.16$ & $-8.31 \pm 0.18$ & $-7.76 \pm 0.16$ \\
 \vspace{0.15cm}
& SN &  $-707.47 \pm 0.20$ & $-707.57 \pm 0.19$ & $-708.01 \pm 0.21$ & $-708.82 \pm 0.23$ & $-706.05 \pm 0.17$\\
 \vspace{0.15cm}
& BAO &  $-12.61 \pm 0.23$ & $-12.57 \pm 0.22$ & $-12.62 \pm 0.23$ & $-13.40 \pm 0.24$ & $-11.69 \pm 0.21$ \\
 \vspace{0.15cm}
 & HD+SN &  $-712.61 \pm 0.23$ & $-712.61 \pm 0.22$ & $-713.22 \pm 0.23$ & $-714.57 \pm 0.25$ & $-711.10 \pm 0.20$\\
 \vspace{0.15cm}
 & BAO+HD &  $-16.01 \pm 0.23$ & $ -16.29 \pm 0.23$ & $-16.21 \pm 0.23$ & $-16.98 \pm 0.25$ & $-15.25 \pm 0.22$ \\
 \vspace{0.15cm}
 & BAO+SN &  $ -715.48 \pm 0.25$ & $-717.12 \pm 0.25$ & $-716.18 \pm 0.26$ & $-717.95 \pm 0.28$ & $-714.81 \pm 0.22$\\
 \vspace{0.15cm}
& BAO+HD+SN &  $-718.81 \pm 0.25$ & $-720.40 \pm 0.25$ & $-719.77 \pm 0.26$ & $-721.90 \pm 0.29$ & $-718.06 \pm 0.22$ \\
\hline
\\
$\ln B_{m,\Lambda}$ & HD &  $-0.99$	& $0.09$ & $0$ & $-0.55$ & $0$ \\
 \vspace{0.15cm}
& SN &  $-1.42$ & $-1.52$ & $-1.96$ & $-2.77$ & $0$\\
 \vspace{0.15cm}
& BAO &  $-0.92$ & $-0.88$ & $-0.93$ & $-1.71$ & $0$ \\
 \vspace{0.15cm}
 & HD+SN &  $-1.51$ & $-1.51$ & $-2.12$ & $-3.47$ & $0$\\
 \vspace{0.15cm}
 & BAO+HD &  $-0.76$ & $-1.04$ & $-0.96$	& $-1.73$ & $0$ \\
 \vspace{0.15cm}
 & BAO+SN &  $-0.67$ & $-2.31$ & $-1.37$ & $-3.14$ & $0$\\
 \vspace{0.15cm}
& BAO+HD+SN &  $-0.75$ & $-2.34$ & $-1.71$ & $-3.84$ & $0$ \\
\hline

\hline
\end{tabular}
\caption{\changes{Constraints at 68\% CL on the parameters, $-2\ln\Like_{\rm max}$ and log-Bayesian 
evidence $\ln \mathcal{Z}$ for quintessence, phantom, quintom and quintom$+\beta$, using 
different datasets.}} 

\label{tab:nestedestimation}
\end{table}


\section{Appendix}\label{appendix_a}

\begin{figure}[h!]
\captionsetup{justification=raggedright,singlelinecheck=false,font=small}
\begin{center}
\includegraphics[trim = 2mm  0mm 0mm 1mm, clip, width=4.cm, height=8cm]{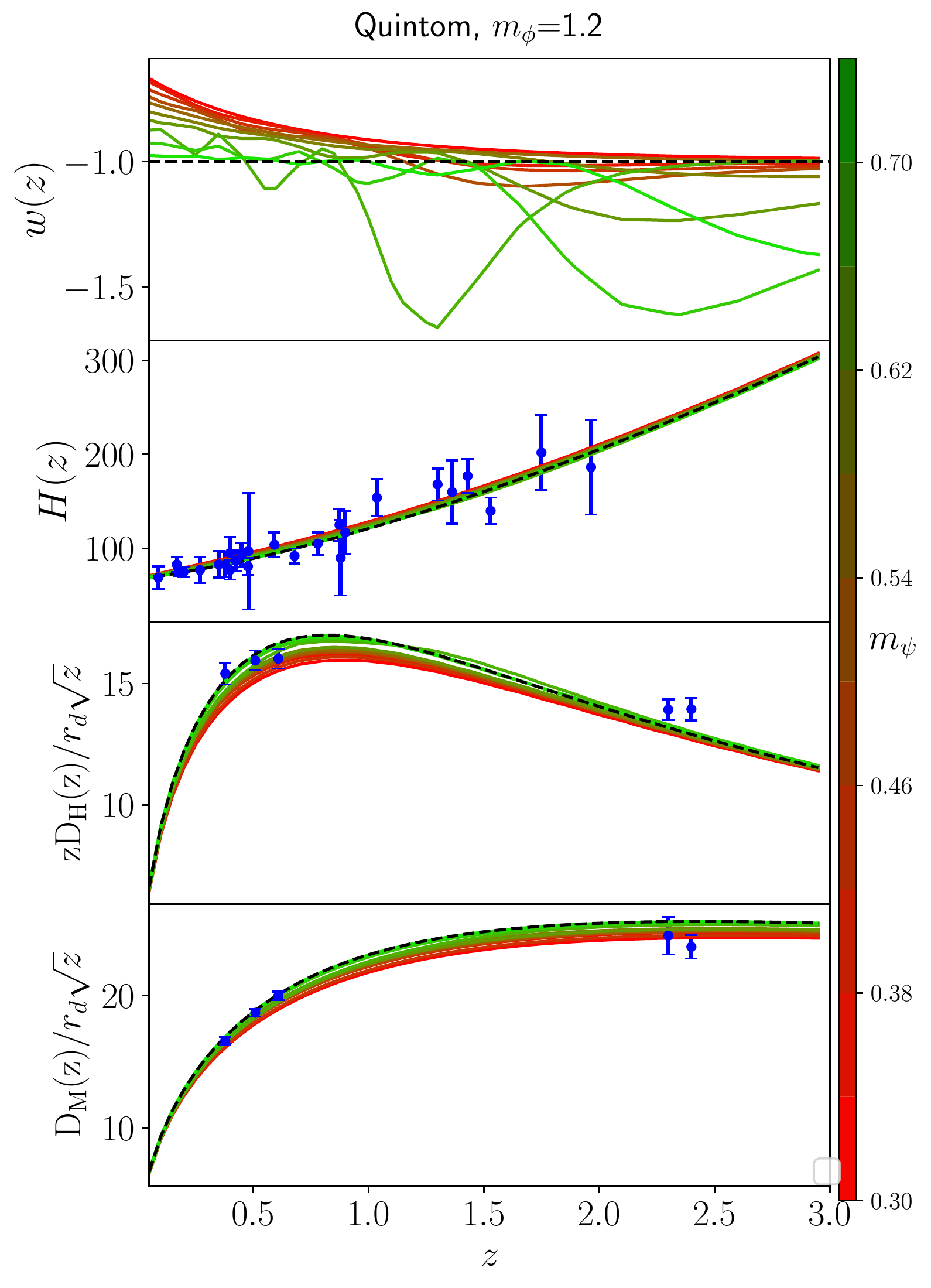} 
\includegraphics[trim = 2mm  0mm 0mm 1mm, clip, width=4.cm, height=8cm]{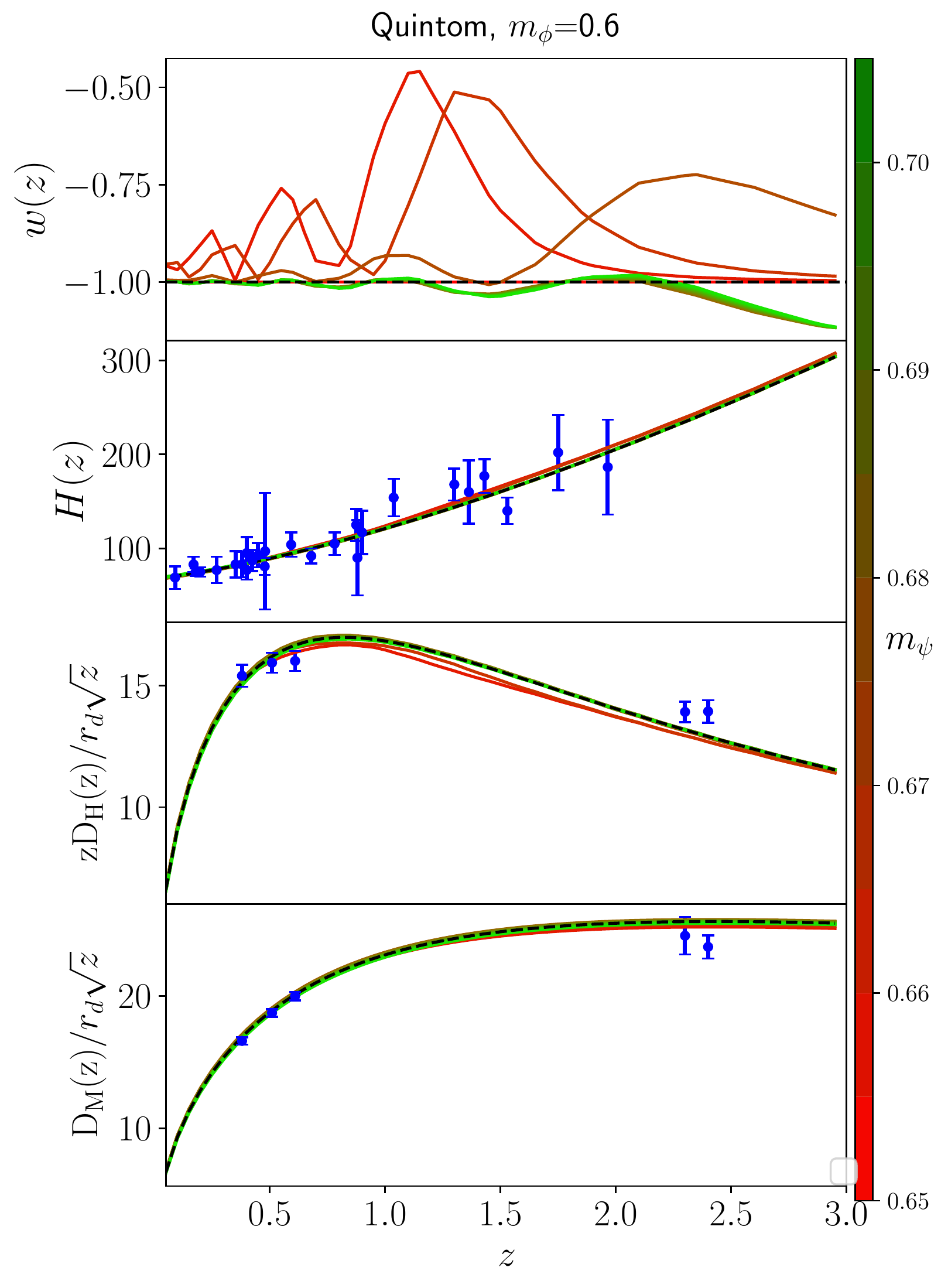} 
\end{center}
\caption[]{Quintom model with oscillatory potential. $w(z)$, $H(z)$, $D_H(z)$ and  $D_M(z)$ for different 
values of the parameters $m_{\phi}$ for the oscillatory potential (\ref{eq:oscillatory}).}
\label{fig:quintom_cos}
\end{figure}

The dynamics of quintom models with various potentials have also been explored using dynamical system techniques. 
One of the studied potentials consists of a sum of individual exponentials, without a direct coupling 
\cite{Guo:2004fq}, and with a particular interaction potential between the phantom and quintessence fields in \cite{Zhang:2005eg}:
\begin{eqnarray}\label{eq:double_exp}
    V &=& V_{\phi 0}\, e^{-\lambda_{\phi}\kappa \phi} + V_{\psi 0}\, e^{- \lambda_{\psi}\kappa\psi}+ V_{\rm int}(\phi, \psi).
    \end{eqnarray}
 For this potential the authors found that there is a future attractor dominated by the phantom field, 
 indicating an EoS below $-1$ at late times. Their analysis also show that, the universe went through distinct 
 stages that allowed the quintessence field to dominate $(w > -1)$ in the past. This implies that the model enables 
 a transition across the PDL irrespective of whether there is a direct coupling between the fields or not. 
 In reference \cite{Guo:2004fq}, it is shown that a similar EoS behavior to the one associated with 
 Eq.(\ref{eq:double_exp}) can be achieved by the Eq.(\ref{eq:Vtot}. 
 However they exhibit that the sum of potentials with a quadratic scalar field in the exponents, given by

 \begin{eqnarray}
    V &=& V_{\phi 0}\, e^{-\lambda_{\phi}\kappa^2 \phi^2} + V_{\psi 0}\, e^{- \lambda_{\psi}\kappa^2 \psi^2},
\end{eqnarray}
leads to an opposite behavior where the EoS transitions from being below $-1$ (at early times) to above $-1$ 
and tends to approach $-1$ at late times, this is similar to our findings for the quintom$+\beta$ with 
$\beta < 0$ (see panel d) of figure \ref{fig:mquin-mphan}).
The EoS can also exhibit an oscillatory  behavior. For instance,
Figure \ref{fig:quintom_cos} shows some cases when varying the mass of the field for the oscillatory potential
\begin{equation} \label{eq:oscillatory}
    V=m_{\phi}^2(1-\cos(a\phi))  + m_{\psi}^2(1-\cos(a\psi)).
\end{equation}
Similar to the behavior of the Eq.(\ref{eq:oscillatory}), in reference \cite{Zhang:2005eg}, the authors 
argue that the following potentials can yield to an oscillatory EoS where the oscillations across $-1$ 
can occur  in the recent past and have the potential to produce observable effects: 
\begin{eqnarray}
    V &=& V_1 \cos\left(\xi \frac{\phi}{M_P}\right) + V_2 \cos\left(\beta \frac{\phi}{M_P}\right) +\beta \phi^2\psi^2,\\
    V &=& \frac{1}{2}m_{\phi}^2\phi^2 +\frac{1}{2}m_{\psi}^2\psi^2 +\beta \psi^2\phi^2.
\end{eqnarray}
Another combination of potentials that has been explored in the quintom scenario is the linear potential, 
\cite{Perivolaropoulos:2004yr, Wu:2005apa}:
\begin{eqnarray}
    V = a(\phi + \psi) + \beta \phi\psi\, .
\end{eqnarray}
By changing the sign of the coupling constant, $\beta$, the model is capable of emulating the distinct 
characteristics exhibited by both quintessence and phantom models, similar to our findings.

\section*{Acknowledgements}
JAV acknowledges the support provided by FOSEC SEP-CONACYT Investigaci\'on B\'asica A1-S-21925, 
UNAM-DGAPA-PAPIIT IN117723 and FORDECYT-PRONACES-CONACYT/304001/2019.
IQ acknowledges FORDECYT-PRONACES-CONACYT for support of the present research under grant CF-MG-2558591. 
AAS acknowledges the funding from SERB, Govt of India under the research grant no: CRG/2020/004347. 
IGV and GGA acknowledges CONAHCYT postdoctoral fellowship and the support of the ICF-UNAM.


\end{document}